\documentclass[aps,prd,onecolumn,superscriptaddress,nofootinbib,floatfix,noshowpacs]{revtex4-1}
\pdfoutput=1
\newif\ifcomment
\usepackage{amsmath,amssymb}
\usepackage{color,hyperref,url}
\usepackage{listings}
\usepackage{slashed}
\usepackage[pdftex]{graphicx}
\usepackage{epstopdf}
\usepackage{epsfig}
\usepackage{grffile}
\usepackage{relsize}
\graphicspath{{./img/}}
\usepackage{soul}
\newcommand{\ga}{\gamma}
\newcommand{\Ga}{\Gamma}

\newcommand{\la}{\lambda}

\newcommand{\beq}{\begin{equation}}
\newcommand{\eeq}{\end{equation}}
\newcommand{\ba}{\begin{array}}
\newcommand{\ea}{\end{array}}
\newcommand{\bea}{\begin{align}}
\newcommand{\eea}{\end{align}}
\newcommand{\bi}{\begin{itemize}}
\newcommand{\ei}{\end{itemize}}
\newcommand{\ben}{\begin{enumerate}}
\newcommand{\een}{\end{enumerate}}
\newcommand{\bc}{\begin{center}}
\newcommand{\ec}{\end{center}}
\newcommand{\bl}{\begin{flushleft}}
\newcommand{\el}{\end{flushleft}}
\newcommand{\br}{\begin{flushright}}
\newcommand{\er}{\end{flushright}}

\newcommand\eqn[1]{(\ref{#1})}      % parentheses around the LaTex "ref" macro
\newcommand\Eqn[1]{Eq.~(\ref{#1})}  % includes ``Eq.'' in front
\newcommand\Fig[1]{Fig.~\ref{#1}} % includes ``Fig.'' in front

\newcommand{\mr}{\mathrm}

\newcommand{\mi}{\mathop{}\!i}

\newcommand{\p}{\partial}

\newcommand{\tr}{\hbox{tr}}

\newcommand{\GeV}{{\rm GeV}}

\renewcommand{\l}{\left}
\renewcommand{\r}{\right}

\renewcommand{\sp}{\shortparallel}

\begin{document}
\title{Exposing the effect of the $p$-wave component in the pion triplet under a strong magnetic field}
\author{Zanbin Xing}%
\email{xingzb@mail.nankai.edu.cn}
\affiliation{School of Physics, Nankai University, Tianjin 300071, China}
\author{Jingyi Chao}%
\email{chaojingyi@jxnu.edu.cn}
\affiliation{College of Physics and Communication Electronics, Jiangxi Normal University, Nanchang 330022, China}
\author{Lei Chang}%
\email{leichang@nankai.edu.cn}
\affiliation{School of Physics, Nankai University, Tianjin 300071, China}
\date{\today}
\author{Yu-xin Liu}\email{yxliu@pku.edu.cn}
\affiliation{Department of Physics and State Key Laboratory of Nuclear Physics and Technology, Peking University, Beijing 100871, China}
\affiliation{Collaborative Innovation Center of Quantum Matter, Beijing 100871, China}
\affiliation{Center for High Energy Physics, Peking University, Beijing 100871, China}
%============================================================
\begin{abstract}
The static properties, masses and decay constants, of a pseudoscalar meson triplet in a strongly magnetized medium are studied through the Dyson-Schwinger equation approach treatment of a contact interaction. Complementary to the usual vector-vector form, a symmetry-preserving formulation of couplings has been proposed in this work, without modifying the quark propagator, to control the strength of the $p$-wave component of Bethe-Salpeter amplitude.  It is found that, with the help of flexible auxiliary interaction, our simple model is able to reproduce the observation in the lattice QCD simulation, where the spectra of the charged pseudoscalar meson shows a nonmonotonic behavior as the magnetic field grows. The results of this work imply the strong magnetic field affects the inner structure of mesons dramatically.
\end{abstract}
\keywords{charged pion, weak decay constant, magnetic fields, $p$ wave}

\maketitle
%============================================================
\section{introduction}\label{sec:int}
A pion, as the QCD's Goldstone boson, plays an essential role in low-energy hadron physics. The deep understanding of a pion's properties in the vacuum and under extreme condition might not only provide insight on the emergent hadron mass but also the phase transition of strong interaction~\cite{Roberts:2021nhw,Klevansky:1992qe,Buballa:2003qv,Brandt:2015sxa,Gao:2020hwo}. Aside from the extensively explored environments at finite temperatures and finite baryon densities, a magnetic field arises as a new dimension to analyze the QCD phases and phase transition~\cite{Agasian2008,D'Elia2010,Mizher2010,Bali:2011qj,Andersen2014}. Motivated by the impact on the evolution of the early universe and the magnetars, where the strength of a magnetic field over its surface is up to $10^{10}\,\mr{T}$~\cite{Thompson2005,Rea:2012ni,kaspi_magnetars_2017}, and boosted by the novel topological phenomena emerging in off-central heavy-ion collisions~ \cite{Skokov2009,Voronyuk2011,deng_event-by-event_2012}, where a stronger $eB$-field, up to $10^{14}-10^{16}\,\mr{T}$, is produced, the magnetized spectra of pions have been examined in various theoretical approaches for low-energy QCD~\cite{Fayazbakhsh:2013cha,Avancini:2015ady,Zhang:2016qrl,Wang2018,Liu2018,Mao2018,GomezDumm2018,Ayala2018,Coppola2019,Sheng:2020hge,Das2020}, within the lattice gauge theory~\cite{Luschevskaya:2015cko} and in the chiral perturbation theory~\cite{Shushpanov:1997sf,andersen_thermal_2012,Colucci:2013zoa}.

When the constituent quarks couple with the magnetic field, the composite neutral pion is disturbed by the external magnetic field as a consequence. However, characterized by the spontaneous chiral symmetry breaking of $U(1)_{I_{3}}\otimes U(1)_{AI_{3}}\to U(1)_{L+R}$~\cite{Chao:2018ejd}, it is expected that the properties of a neutral pion are preserved by the Goldstone nature. Presented by the research groups of~\cite{Bali2018,Ding2020}, the lattice QCD (LQCD) computations conclude that $\pi_{0}$ is as light as a pseudo-Goldstone boson. Moreover, an interesting observation is that the mass of neutral pions monotonically decreases as a function of the field strength. Some discussions from low-energy approximations of QCD are given in Ref.~\cite{Xu:2020yag}. On the contrary, these two nonperturbative calculations contradict each other on the energy dispersion of charged pions. As a tight bound state, the spectra of $\pi_{\pm}$ are perfectly amenable to the formula of $E^{2}(n,p_{z})=p_{z}^2+(2n+1)eB+m_{0}^{2}$ at a weak limit for $eB\leq 0.4\,\GeV^2$. Compared to $\pi_{0}$, the behaviors of charged mesons are more sensitive to the underlying dynamic of QCD$\otimes$QED while turning on $eB$ and its point-particle approximation must become corrupt in a large region of the $eB$ field. However, as found in Ref.~\cite{Ding2020}, the masses of charged pions and kaons linearly increase at the beginning, grow slowly, reach the greatest point, decrease as the magnetic field increases, and end up with a saturation at $eB \gtrsim  2.5 \text{GeV}^2$. Such a nonmonotonic mass curve of charged pions in terms of the magnetic field is beyond the prediction of low-energy effective theories~\cite{andersen_thermal_2012,Colucci:2013zoa,Luschevskaya:2015cko,Zhang:2016qrl,Wang2018,Liu2018,Mao2018,Coppola2019}. Meanwhile, Ref.~\cite{Bali:2011qj} shows that, after a same uniform increase and a similarly slower growing, the mass of charged mesons saturates directly in the ultrastrong limit. Thus, several theoretical efforts have been made (see, e.g., Refs~\cite{Xu:2020yag}) to address the discrepancy between the two first principle results. 

Although the pseudoscalar meson is at the state of $J=0^{-+}$, its internal structure contains a nontrivial $p$-wave component, 
which has been discussed in the Bethe-Salpeter equation (BSE) formalism (see, e.g., Refs.~\cite{Hilger:2015ora,Bhagwat:2006xi}). With regard to this concern, an intuitive conjecture is that if the higher angular momentum plays an important role in equipping the Fock states of magnetized pions, it will naturally reduce the mass of charged mesons. In other words, the isovector meson's zero-point energy arising in magnetic fields would be dismissed by the total spin of two valence quarks like the magnetized vector meson~\cite{Chernodub:2010qx}. To demonstrate this idea, in the present paper, we will employ the Dyson-Schwinger equations (DSE) to study the pions and their properties taking the advantage that the BSE-DSE includes intrinsically the context of $p$ waves in pseudoscalar mesons. As an early investigation, the contact interaction is adopted. Hence, in order to flexibly control the strength of the $p$ wave, we blend in an auxiliary term in the interactions, which only manifests itself in the meson sector. In turn, such an alternative method provides an efficient way to tune the magnitudes of a meson's scattering kernels.

Furthermore, the scope of this paper will extend to the analysis of the pion-to-vacuum matrix elements of the axial vector hadronic current based on the same model approach. Taking into account the energy-momentum tensors of the magnetic field and numerating all the allowed Lorentz structures, three leptonic decay constants of charged pions arose. The conventional weak decay constant $f_{\pi}$ has been analyzed by different points of view~\cite{bali_weak_2018} and another emerging two have been achieved in the approach of the Nambu--Jona-Lasinio (NJL) model model~\cite{Coppola2019}. We note here that we will evaluate only two decay constants as the functions of the magnetic field, since one of them will vanish in the presented contact model.

This remainder of this paper is organized as follows. In Sec.~\ref{sec:oringal}, we describe briefly the full formalism in the sectors of quarks and mesons under the influence of a uniform magnetic field background and compute the quark masses, pion masses, and weak decay constants with the conventional vector-vector contact interactions. In Sec.~\ref{sec:model}, we concentrate on the role of the $p$ wave in the meson sector and its impact on the properties of pions and repeat the calculations containing the auxiliary term. Our conclusions are summarized in Sec.~\ref{sec:con}.

\section{general formula in the conventional contact model}\label{sec:oringal}
\subsection{Quark sector}
The strong interaction at low-energy scale is completely nonperturbative and extremely difficult by itself, and becomes harder to deal with in the presence of magnetic fields. In this section, we review the model and the formalism applied in our calculations.

To account for the minimal electromagnetic (EM) coupling along the $z$ direction, the Landau gauge potential $A_{\mu}=\l(0,Bx,0,0\r)$ is used. For further simplification, a vector-vector contact interaction in coordinate space is introduced to mimic the strong coupling~\cite{GutierrezGuerrero2010,Roberts2011},
\begin{equation}
\frac{4}{3 m_G^2}\delta^{4}(a-b)\gamma_{\mu}\otimes\gamma_{\mu},
\end{equation}
where $a,b$ denote the coordinate $\l(x,y,z,t\r)$. The mass scale of gluon $m_G$ stands for the strength of the instantaneous interaction and is fixed in order to fit the empirical values. Implementing the contact interaction, the quark gap equation  develops to
\begin{equation}\label{gapeq}
\begin{aligned}
S^{-1}\!\left(a,b\right)=S_0^{-1}\!\left(a,b\right)+\frac{4}{3 m_G^2}\delta^4\!\left(a-b\right)\!\gamma_\mu S(a,b)\gamma_\mu,
\end{aligned}
\end{equation}
where the inverse of the bare quark propagator
\begin{equation}
S_0^{-1}\!\left(a,b\right)=\delta^{4}\left(a-b\right)\left(\slashed{\partial}_b+ie_f\slashed{A}+m_0\right),
\end{equation}
and $m_{0}$ is the current quark mass, uniform in the two-flavor space.
Besides, a general structure of the inverse quark propagator is described as
\begin{equation}\label{inversedressedpropagator}
S^{-1}(a,b)=\delta^{4}(a-b)\left(\slashed{\partial}_b+ie_f\slashed{A}+M_f\right),
\end{equation}
where the dynamical quark mass $M_f$ is a static constant with $f=u,d$.
To obtain the  propagator, the usual way is to construct a complete orthogonal basis and then to form the propagator through it.
Belonging to the different eigenvalues of the Hermitian operator, the magnetized eigenfunctions carry out a complete set, and it allows us to apply the operator $G^{\prime\,2}\left(a,b\right)$ with $G^{\prime}\left(a,b\right)=\gamma_{5}S^{-1}\left(a,b\right)$, which are governed by the eigenequations
\begin{equation}\label{eigeneq}
\int d^4b G^{\prime\,2}\left(a,b\right)f\left(b\right)=\lambda^2f\left(a\right).
\end{equation}
It is observed that $f(a)$ follows the equation of motion of the simple harmonic oscillator and the corresponding eigenvalue is double degenerate excluded at the lowest Landau level (LLL).

Employing the basis derived from Eq.~(\ref{eigeneq}), the propagator in $(x,q_y,q_z,q_t)$ space (we call it as a representation in momentum space in the rest of the paper, even coordinate $x$ remains) is written as
\begin{equation}\label{propagator}
\begin{aligned}
S\left(\bar{q};x,x^\prime\right)=\sum_{n=0}^{\infty}
&\left\{P_{+}^{\xi}\frac{-i\slashed{q}+M_f}{q^2+M_n^2}f_n\left(x,q_y\right)f_n\left(x^\prime,q_y\right)-I_{+}^{\xi}\frac{\sqrt{2B_{n}}}{q^2+M_n^2}f_{n-1}\left(x,q_y\right)f_n\left(x^\prime,q_y\right)\right.\\
&\left.+I_{-}^{\xi}\frac{\sqrt{2B_{n+1}}}{q^2+M_{n+1}^2}f_{n+1}\left(x,q_y\right)f_n\left(x^\prime,q_y\right)+P_{-}^{\xi}\frac{-i\slashed{q}+M_f}{q^2+M_{n+1}^2}f_n\left(x,q_y\right)f_n\left(x^\prime,q_y\right)\right\},
\end{aligned}
\end{equation}
where the incomplete-dimension vectors read $\bar{q}=(q_y,q_z,q_t)$ and $q=(q_z,q_t)$, 
and the full four-dimension vectors are $\bar{q}_\mu=(0,q_y,q_z,q_t)$ and $q_\mu=(0,0,q_z,q_t)$. Note that $\xi=\mr{sign}(e_fB)$, $P_{\pm}^{\xi}=\frac{1\pm i\xi\gamma_1\gamma_2}{2}$, $I_{\pm}^{\xi}=\frac{\gamma_1\pm i\xi\gamma_2}{2}$, $B_{n}=n\left|e_fB\right|$, $M_{n}^{2}={M_f^2+2B_{n}}$, and $f_n\left(x,p_y\right)=\hat{f}_n\left(x+\frac{p_y}{e_fB}\right)$. The normalized harmonic oscillator wave functions $\hat{f}_n\left(x\right)$ take the form of
\begin{equation}\label{nhwf}
\begin{aligned}
\hat{f}_n\left(x\right)=\frac{\left|e_fB\right|^{1/4}}{\sqrt{2^{n}n!\sqrt{\pi}}}H_n\left(\sqrt{\left|e_fB\right|}x\right) e^{-\left|e_fB\right|x^2/2},
\end{aligned}
\end{equation}
where $H_n\left(x\right)$ are the Hermitian polynomials. A detailed formalism of the propagator $\lim_{a\to a'}S(a,a')$ can be found in Appendix~\ref{App:quark}.

\begin{figure}[ht]
\includegraphics[width=0.5\textwidth]{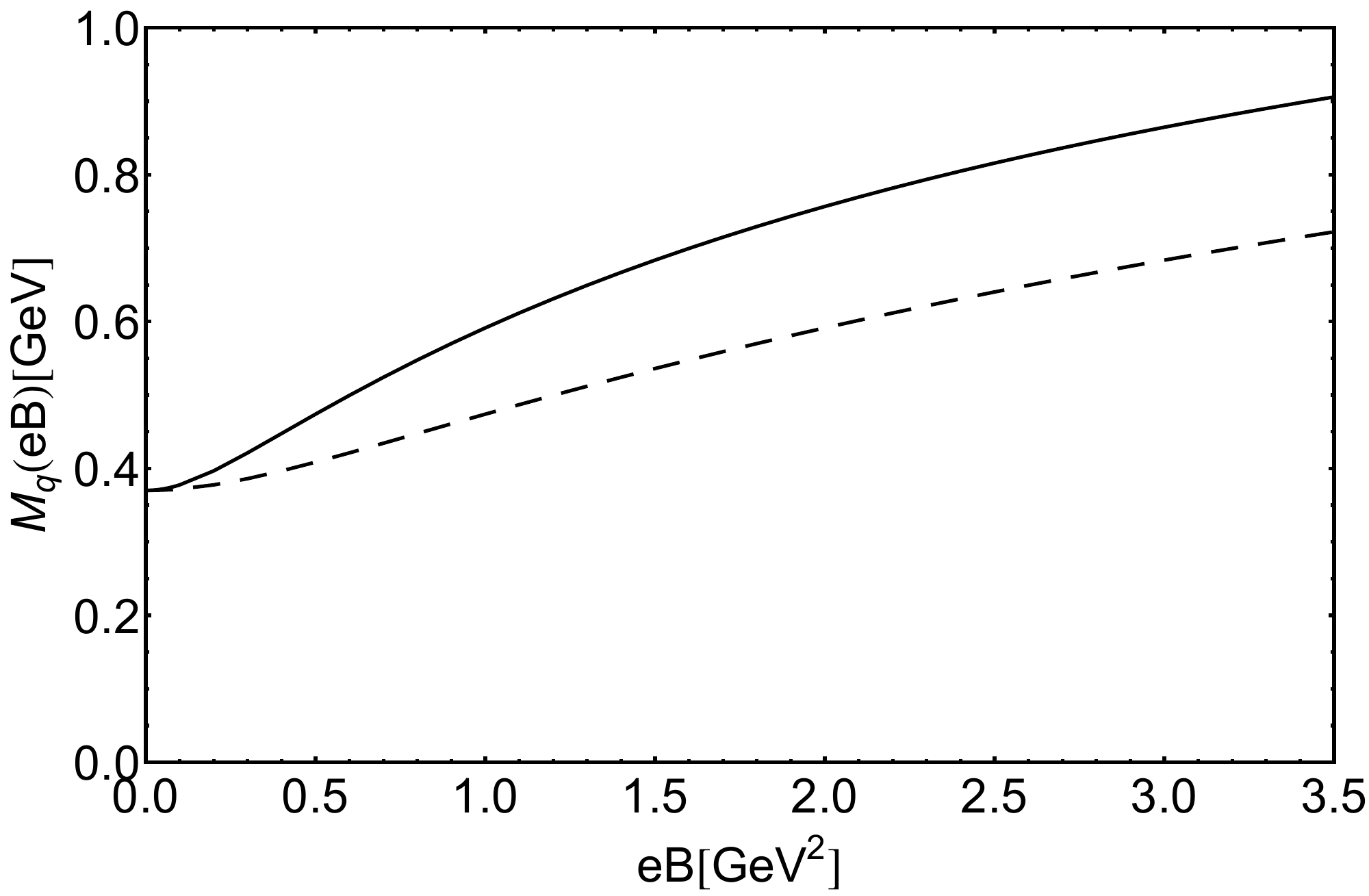}
\caption{Calculated quark masses as the function of magnetic fields in contact interaction. The solid line denotes the dynamic mass of a $u$ quark and the dashed line represents the mass of a $d$ quark. }\label{quark result}
\end{figure}

Plugging in the magnetized quark propagator, the gap equation of the dynamical mass transforms to
\begin{equation}\label{mass}
\begin{aligned}
M_f=m_0+\frac{8M_f}{3 m_G^2}\int_{-\infty}^{\infty}\frac{d^2q}{\left(2\pi\right)^2}\sum_{n=0}^{\infty}\frac{\left|e_fB\right|}{2\pi}\left[\frac{1}{q^2+M_n^2}+\frac{1}{q^2+M_{n+1}^2}\right].
\end{aligned}
\end{equation}
The regarded integration and summation are apparently ultraviolet (UV) divergent. Thus, a regularization scheme must be adopted. In this paper, we employ the scheme of the proper time regularization~\cite{GutierrezGuerrero2010}, which is applied as
\begin{equation}
\begin{aligned}
\!\!\!\frac{1}{s+M^2}=\int_0^\infty d\tau e^{-\tau\left(s+M^2\right)}\rightarrow\int_{\tau_{uv}^2}^{\tau_{ir}^2} d\tau e^{-\tau\left(s+M^2\right)}.
\end{aligned}
\end{equation}
The UV parameter $\tau_{uv}$ characterizes the dynamical energy scale of all dimensional quantities in the contact model while as the infrared parameter $\tau_{ir}$ remains as the description of confinement.
The whole used parameters of the model are
\begin{equation}\label{para}
m_G=0.132\,\GeV,\ \tau_{uv}=1/0.905\,\GeV^{-1},\ \tau_{ir}=1/0.24\,\GeV^{-1},\ m_0=0.0089\,\GeV,
\end{equation}
which gives $M_q=0.37\,\GeV$ in vacuum. In the energy scale larger than $eB=0.1~\text{GeV}^2$, the Landau level is chosen to be truncated at $n=50$. To compute in the weak and moderate magnetic fields, we analytically address the infinite summation of Landau levels by the hyperbolic functions, seen the \Eqn{eqn:summedgap} in the Appendix. After summation, we keep applying the above UV and IR truncations with respect to the integration of the proper time $\tau$. We note here such choice of regularization, which treats the magnetic and nonmagnetic contributions the same, is named as nonmagnetic field independent regularization scheme~\cite{Avancini:2019wed}.
The numerical results in \Fig{quark result} demonstrate that the up/down quark mass increases with magnetic field and the quark masses satisfy a charge relation of $M_u(eB)=M_d(2eB)$. One can also see that the mass plotting presents the well-known magnetic catalysis, which is produced by the first term of the right-hand side of Eq.~(\ref{mass}).
%%%%%%%%%%%%%%%%%%%%%%%%
%%%%%%%%%%%%%%%%%%%%%%%%
\subsection{Meson sector}
Due to the coupling between the magnetic field and charge particles, the Bethe-Salpeter equation is no longer translation invariant and will be written in coordinate space. The ladder truncation entails it taking the form
\begin{equation}
\begin{aligned}
\Gamma\left(b,b'\right)=-\int_{-\infty}^{\infty} d^4a\,d^4a'\,\gamma_\mu S_1\left(b,a\right)\Gamma\left(a,a'\right)S_2\left(a',b\right)\gamma_\nu D_{\mu\nu}(b,b'),
\end{aligned}
\end{equation}
where $D_{\mu\nu}(b,b')=\frac{4\delta_{\mu\nu}}{3 m_G^2}\delta^4(b-b')$. Also, the employed Bethe-Salpeter amplitude (BSA) is $\Gamma\left(b,b'\right)=\Gamma\left(b\right)\delta^4(b-b')$, where $\delta^4(b-b')$ implies a pointlike inner structure of composite particles and $\Gamma\left(b\right)$ describes the motion of composite degree of freedoms as a whole. Then, the BSE becomes
\begin{equation}\label{BSEC}
\begin{aligned}
\Gamma\left(b\right)=
-\frac{4}{3 m_G^2}\int_{-\infty}^{\infty} d^4a\,\gamma_\mu S_1\left(b,a\right)\Gamma\left(a\right)S_2\left(a,b\right)\gamma_\mu.
\end{aligned}
\end{equation}
Converting to momentum space, the BSE is written as
\begin{equation}\label{BSE}
\begin{aligned}
\Gamma\left(\bar{P};x\right)=-\frac{4}{3 m_G^2}\int_{\bar{q},x'}\!\gamma_\mu S_1\left(\bar{q},x,x'\right)\Gamma\left(\bar{P};x'\right)S_2\left(\bar{k},x',x\right)\gamma_\mu,
\end{aligned}
\end{equation}
where $\int_{\bar{q},x'}=\int_{-\infty}^{\infty}\frac{d^3\bar{q}}{\left(2\pi\right)^3}\int_{-\infty}^{\infty}dx'$, $\bar{k}=\bar{q}-\bar{P}$. 
In general, one has $S\left(x,x'\right)=\phi\left(x,x'\right)\tilde{S}\left(\bar{q},x-x'\right)$ (see, e.g., Ref.~\cite{Kojo:2013uua}), where the position dependence of the magnetic field is attributed to the Schwinger phase factor $\phi\left(x,x'\right)$ and the left $\tilde{S}$ is translation invariant. Those contributions from the phase factor are trivial for mesons, since only two propagators are involved in \Eqn{BSE} (a more detailed explanation can be found in Ref.~\cite{Chyi:1999fc}).
Hence, Fourier transforming the translation invariant part and considering the screening mass along the $z$ direction, one is allowed to use the incomplete-dimension momentum $\bar{P}=\left(0,P_z,0\right)$. To stress, the homogeneous BSE Eq.~(\ref{BSE}) is valid only when the on-shell condition $\bar{P}^2=-m_\pi^2$ is satisfied. To numerically find out the value of $m_\pi^2$ in a more efficient way, we transform the above equation into an eigenvalue problem in the form
\begin{equation}\label{BSEeigen}
\begin{aligned}
\lambda(\bar{P}^2)\Gamma\left(\bar{P};x\right)=-\frac{4}{3 m_G^2}\int_{\bar{q},x'}\!\gamma_\mu S_1\left(\bar{q},x,x'\right)\Gamma\left(\bar{P};x'\right)S_2\left(\bar{k},x',x\right)\gamma_\mu,
\end{aligned}
\end{equation}
where $\lambda(\bar{P}^2)$ is the eigenvalue. One notices that $\la$ is a function of $\bar{P}^2$, and its magnitude is equivalent to the appearance of meson poles while
\begin{equation}\label{massshell}
\lambda(\bar{P}^2=-m_\pi^2)=1.
\end{equation}

In the contact model, the general structures of pseudoscalar meson's BSA are given as
\begin{equation}\label{BSA}
\Gamma_{\pi}\left(\bar{P};x\right)=\gamma_5 \left(E_{\pi}\left(\bar{P};x\right)+\frac{-i\slashed{\bar{P}}}{2M}F_{\pi}\left(\bar{P};x\right)\right),
\end{equation}
where $M=\frac{M_{1}M_{2}}{M_{1}+M_{2}}$, with $M_{1,2}$ being the effective mass of propagator $S_{1,2}$. $E$ and $F$ characterize the corresponding pseudoscalar and axial-vector components, respectively.
Without a magnetic field, the pion mass and leptonic decay constant are $m_{\pi}(eB=0)=0.157\,\GeV$, $f_{\pi}(eB=0)=0.102\,\GeV$ restricted with both finite $E$ and $F$ terms, while as $m_{\pi}(eB=0)=0.135\,\GeV$, $f_{\pi}(eB=0)=0.119\,\GeV$ with a single $E$ component (i.e.~setting $F=0$) in the BSA.

\subsubsection{Neutral pion}
The neutral pion state $\left|\pi_0\right>=\alpha(B)\left|\pi_u\right>-\beta(B)\left|\pi_d\right>$ is a mixture of $u,d$ quarks. The associated coefficients satisfy $\alpha^2(B)+\beta^2(B)=1$ and $\alpha(B)=\beta(B)\rightarrow\frac{1}{\sqrt{2}}$ for $B\rightarrow0$~\cite{Bali2018}. 
Strictly speaking, the coefficients should be determined by other quantities which relate to experiment observations of $\pi_0$. However, to keep our discussion simpler, we compute the pure states of $\left|\pi_u\right>$ and $\left|\pi_d\right>$ and estimate the real pion mass located between them. Here, we will exhibit $\pi_0^u$ as an example.
We assume that the Bethe-Salpeter amplitude for neutral particles is a delta function in coordinate space, i.e.~$E_{\pi_0^u}\left(\bar{P};x\right)=E\delta(x)$ with $E$ being a constant and so does $F_{\pi_0}\left(\bar{P};x\right)$. Since the integration with respect to $q_y$, $x$, and $x'$ can be done analytically, it reduces the kernels to
\begin{equation}\label{neutralpionBSE}
\begin{aligned}
\left[\begin{array}{c}E\\F\end{array}\right]=\frac{4}{3 m_G^2}\left[\begin{array}{cc}K_{EE}^{\pi_{0}^u}&K_{EF}^{\pi_{0}^u}\\K_{FE}^{\pi_{0}^u}&K_{FF}^{\pi_{0}^u}\end{array}\right]\left[\begin{array}{c}E\\F\end{array}\right],
\end{aligned}
\end{equation}
where the kernel $K_{EE}^{\pi_{0}^u}$ is
\begin{equation}
\begin{aligned}
K_{EE}^{\pi_{0}^u}&=-\tr\int_{\bar{q},x}\frac{\gamma_5}{4}\gamma_\mu S_u\left(\bar{q},x,0\right)\gamma_5S_u\left(\bar{k},0,x\right)\gamma_\mu\\
&=\tr\int_{\bar{q},x}{\gamma_5S_u\left(\bar{q},x,0\right)\gamma_5S_u}\left(\bar{k},0,x\right)\\
&=\frac{\left|e_uB\right|}{2\pi}2\sum^\infty_{n_u=0}G_{1,n_u}+G_{2,n_u}+G_{3,n_u}+G_{4,n_u}.
\end{aligned}
\end{equation}
The explicit forms of $G_{j,n_u}$ and the other three kernels are listed in Appendix~\ref{App:BSE}.

To manifest the Goldstone nature of a neutral pseudoscalar meson in chiral limit $m_{0}=0$, one writes down the first two elements of the matrix Eq.~(\ref{neutralpionBSE}), shown as
\begin{equation}
\begin{aligned}
&K_{EE}^{\pi_{0}^u}=\frac{4}{3m_G^2}\int\frac{d^2q}{\left(2\pi\right)^2}\sum^{\infty}_{n_u=0}\frac{\left|e_uB\right|}{2\pi}\left\{\frac{2}{q^2+M_{n_u}^2}+\frac{2}{q^2+M_{n_u+1}^2}\right\},\quad K_{EF}^{\pi_{0}^u}=0.
\end{aligned}
\end{equation}
Combined with the gap equation of Eq.~(\ref{mass}), it is easy to realize that the demanded solution of Eq.~(\ref{massshell}) has been fulfilled automatically by $m_{\pi}^2=-\bar{P}^2=0$. Moreover, any mixed state in a two-flavor space of magnetized neutral pion is massless as required. 
To numerically compute the pion mass as a pseudo-Goldstone boson, the scheme of regularization is applied in the BSE, as well, which is formulated as 
\begin{equation}
\int\frac{d^2q}{(2\pi)^2}\frac{q\cdot k+A}{\left(q^2+B\right)\left(k^2+C\right)}\ \quad\text{or}\quad \int\frac{d^2q}{(2\pi)^2}\frac{q\cdot P\ A-k\cdot P\ A'}{\left(q^2+B\right)\left(k^2+C\right)},
\end{equation} 
where $A$, $A'$, $B$, and $C$ are irrelevant to either $q$ or $k=q-P$. The UV and IR truncation in momentum space are adopted as
\begin{equation}\label{reg1}
\begin{aligned}
&\int\frac{d^2q}{(2\pi)^2}\frac{q\cdot k+A}{\left(q^2+B\right)\left(k^2+C\right)}=\int_0^1d\alpha\left[\alpha\left(\alpha-1\right)P^2+A\right]{\bar{D}}_1^{iu}\left[\varsigma\left(\alpha,P^2,B,C\right)\right]+D^{iu}\left[\varsigma\left(\alpha,P^2,B,C\right)\right]-D_1^{iu}\left[\varsigma\left(\alpha,P^2,B,C\right)\right],\\
&\int\frac{d^2q}{(2\pi)^2}\frac{q\cdot P\ A-k\cdot P\ A'}{\left(q^2+B\right)\left(k^2+C\right)}=P^2\int_0^1d\alpha\left[\alpha\ A-\left(\alpha-1\right)\ A'\right]{\bar{D}}_1^{iu}\left[\varsigma\left(\alpha,P^2,B,C\right)\right],
\end{aligned}
\end{equation}
where 
\begin{equation}\label{reg2}
\begin{aligned}
&D^{iu}\left[\varsigma\right]=\int_{\tau_{uv}^2}^{\tau_{ir}^2}d\tau\int\frac{d^2q}{(2\pi)^2}e^{-\tau\left(q^2+\varsigma\right)},
\ {\bar{D}}_1^{iu}\left[\varsigma\right]=-\frac{d}{d\varsigma}D^{iu}\left[\varsigma\right],
\ D_1^{iu}\left[\varsigma\right]=\varsigma{\bar{D}}_1^{iu}\left[\varsigma\right],\\
&\varsigma\left(\alpha,P^2,B,C\right)=\alpha \left(1-\alpha\right)P^2+\left(1-\alpha\right)B+\alpha C.
\end{aligned}
\end{equation}

\begin{figure}[ht]
\includegraphics[width=0.5\textwidth]{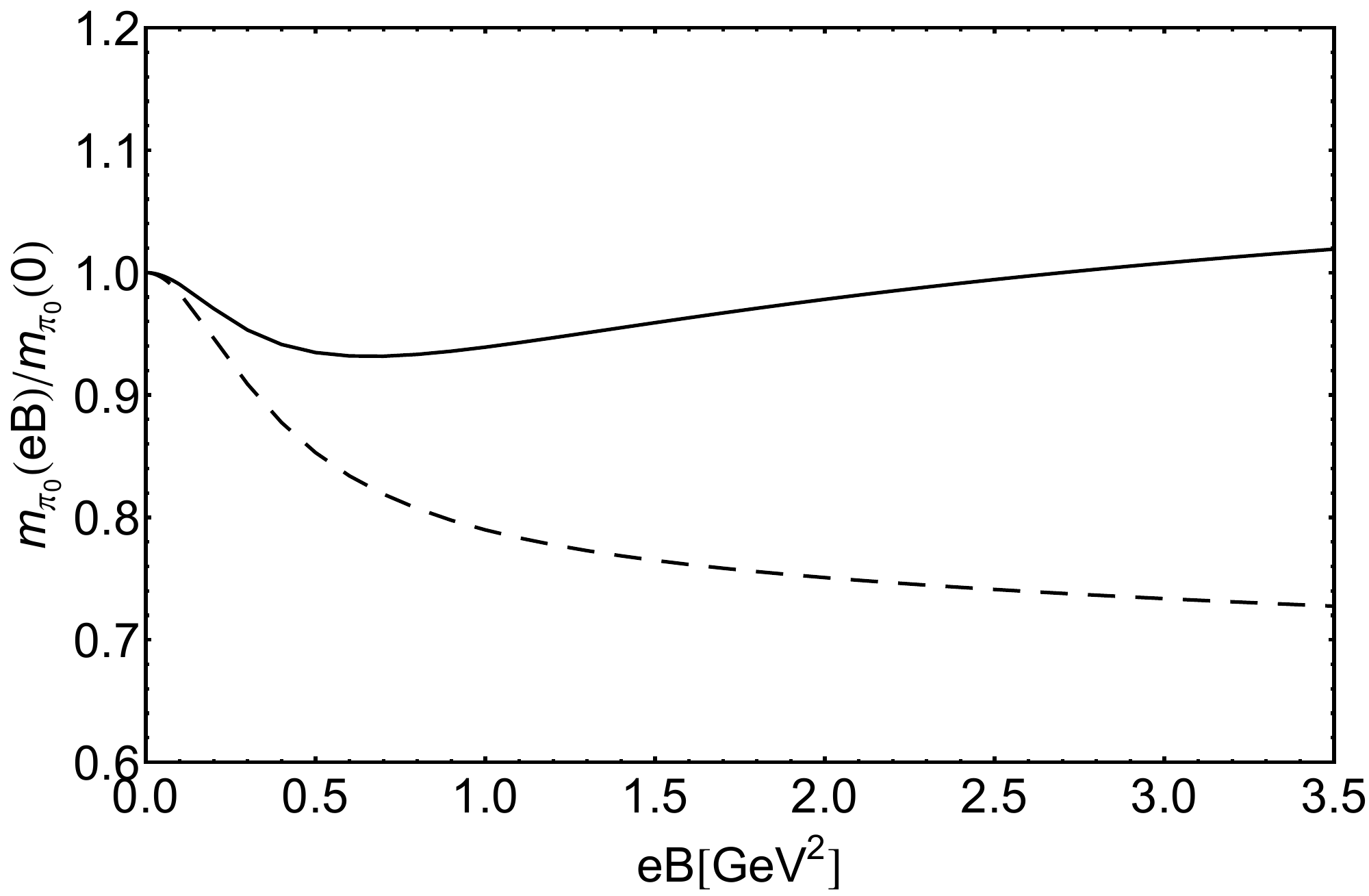}
\caption{Calculated results of the mass of neutral pion $m_{\pi_{0}^{u}}$ as the function of the magnetic field in the conventional contact interaction model. The solid line denotes the mass with complete BSA incorporating both $E$ and $F$; the dashed line denotes the result extracted from an individual $E$ term in the BSA.}\label{neutral pion mass result}
\end{figure}

The obtained neutral pion masses are shown in \Fig{neutral pion mass result}. While the BSA incorporates both $E$ and $F$, the mass first decreases and then increases as the magnetic field grows.
In the context of a single $E$, the mass first decreases but then saturates. Obviously, the component of $F$ plays a non-negligible role. It is also observed that an individual $E$ term in the BSA renders a better result to accommodate with the computations of LQCD~\cite{Ding2020,Bali2018}, which is artificial and a subtler consideration will show in the next section.
%%%%%%%%%%%%%%%%
%%%%%%%%%%%%%%%%
\subsubsection{Charged pion}
Unlike the neutral pion, $\pi_{\pm}$ are no longer protected by the Goldstone theorem and become massive under the influence of external magnetic fields. Also, the Bethe-Salpeter amplitude is not as simple as a delta function but formed by a more complicated Gaussian-like structure. For $\pi_+$, the constituent quarks are $u\bar{d}$ and the Bethe-Salpeter equation is
\begin{equation}\label{chargedpionBSE}
\begin{aligned}
\left[\begin{array}{c}E_{\pi_{+}}\\F_{\pi_{+}}\end{array}\right]=\frac{4}{3 m_G^2}\int dx^\prime \left[\begin{array}{cc}K_{EE'}^{\pi_{+}}&K_{EF'}^{\pi_{+}}\\K_{FE'}^{\pi_{+}}&K_{FF'}^{\pi_{+}}\end{array}\right]\left[\begin{array}{c}E'_{\pi_{+}}\\F'_{\pi_{+}}\end{array}\right],
\end{aligned}
\end{equation}
where $(E,F)$ and $(E',F')$ are functions of $\left(\bar{P};x\right)$ and $\left(\bar{P};x'\right)$, respectively. The detailed form of kernel $K_{EE'}^{\pi_{+}}$ is
\begin{equation}\label{kernel}
\begin{aligned}
K_{EE'}^{\pi_{+}}&=-\tr\frac{\gamma_5}{4}\int_{\bar{q}}\gamma_\mu S_u\left(\bar{q},x,x^\prime\right)\gamma_5S_d\left(\bar{k},x^\prime,x\right)\gamma_\mu\\
&=\tr\int_{\bar{q}}{\gamma_5S_u\left(\bar{q},x,x^\prime\right)\gamma_5S_d}\left(\bar{k},x^\prime,x\right)\\
&=2\sum^{\infty}_{n_u=0}\sum^{\infty}_{n_d=0}G_{1,n_u,n_d}g_{1,n_u,n_d}+G_{2,n_u,n_d}g_{2,n_u,n_d}+G_{3,n_u,n_d}g_{3,n_u,n_d}+G_{4,n_u,n_d}g_{4,n_u,n_d}.
\end{aligned}
\end{equation}
Here $G_{j,n_u,n_d}$ and $g_{j,n_u,n_d}$ have integrated with respect to $(q_z,q_t)$ and $q_y$, respectively. 
The coordinate dependence of $x$ and $x'$ remains in $g_{j,n_u,n_d}$. The UV regularization scheme has been applied to obtain the functions of $G_{j,n_u,n_d}$, and their explicit forms are shown in Appendix~\ref{App:BSE}. 

As shown in \Fig{charged pion mass result}, for BSA incorporating both $E$ and $F$, $\pi_{+}$ mass increases rapidly and the solution of the mesonic bound state disappears for $eB>0.8\,\mr{GeV}^2$ in the traditional model of contact interaction. Such strong enhancement of mass and the disappearance of the bound state are produced by the incorrect ansatz of the $F$ term. Indeed, for a single $E$ structure adopted in the BSA, the spectra of $\pi_{+}$ is very close to the result of the LLL approximation. As we mentioned before, a careful analysis of the $F$ component is necessary and will be addressed in the next section.
%%%%%%%%%%%%%%%%%%%%%%%%%%%%%%
\begin{figure}[ht]
\includegraphics[width=0.5\textwidth]{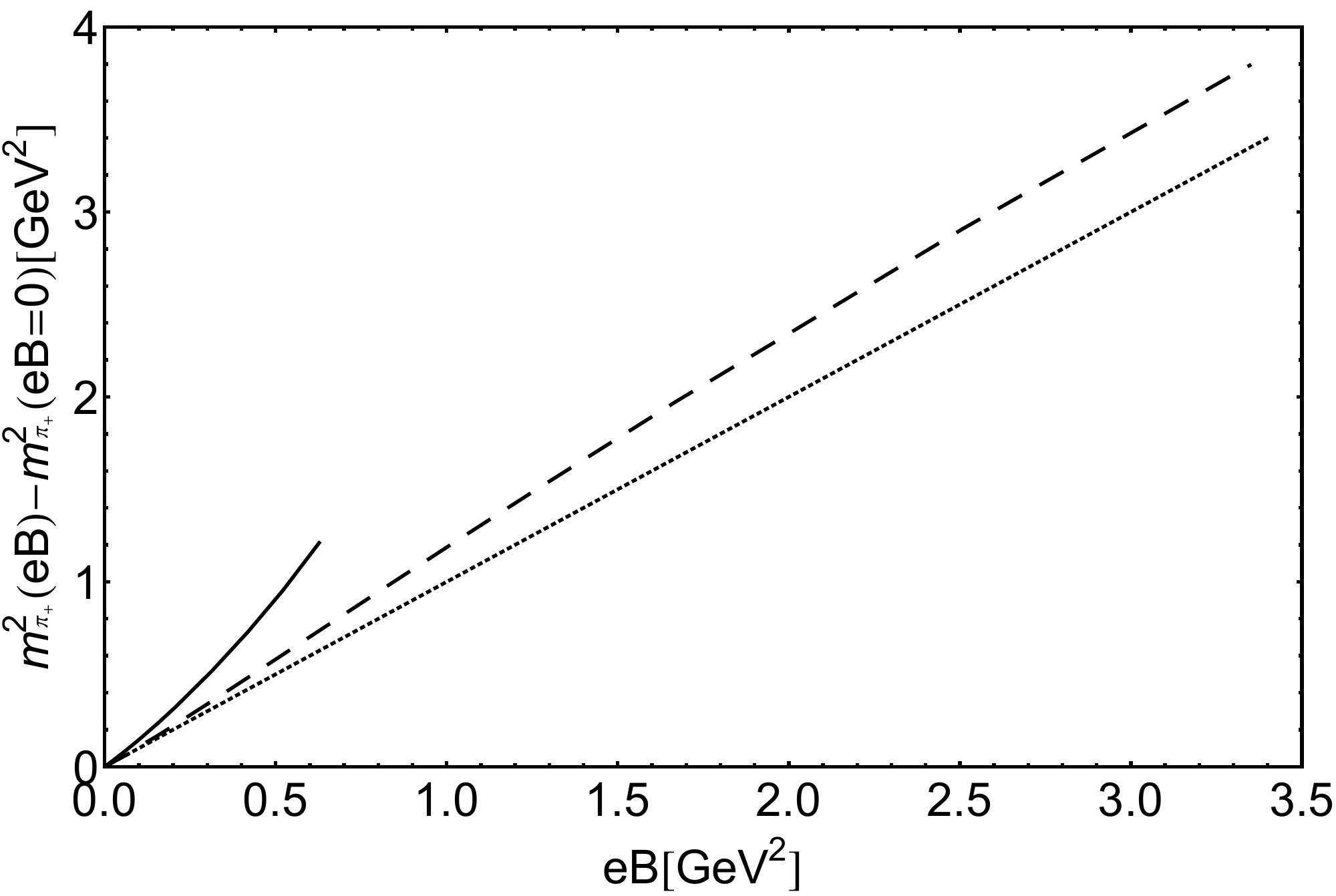}
\caption{Calculated results of the charged pion mass as the function of the magnetic field in the conventional contact interaction model. 
The solid line denotes the mass of $\pi_{+}$ with complete BSA incorporating both $E$ and $F$;
the dashed line denotes the result extracted from an individual $E$ term in the BSA,
and the dotted line represents the LLL approximation.}\label{charged pion mass result}
\end{figure}
%%%%%%%%%%%%%%%%%%%%%%%%%%%%%%
\subsection{Weak decay constants}
According to the equivalence theorem, the Yukawa coupling in pseudoscalar channels can be rewritten to pseudovector types via the integration by part:
\begin{equation}
    \pi\bar{\psi}\ga_5\psi\to g^{\mu\nu}\p_{\mu}\pi\,J_{5\nu},
\end{equation}
where the axial current $J_{5\nu}=\bar{\psi}\ga_5\ga_\nu\psi$. Within the EM field, as a consequence, two more weak decay constants are required to describe the one-pion-to-vacuum matrix elements by virtue of the couplings of $\l(F,\tilde{F}\r)^{\mu\nu}\p_{\mu}\pi\,J_{5\nu}$. Here, those three decay constants $f_{\pi},\ f'_{\pi},\ f''_{\pi}$ are defined as
\begin{equation}\label{decay1}
f_{\pi}\bar{P}_{\mu}+f''_{\pi}eF_{\mu\nu}\bar{P}^{\nu}=N_ctr_D\int_{x,x',\bar{q}}i\gamma_5\gamma_\mu S(\bar{q};x,x')\Gamma_{\pi}(\bar{P};x')S(\bar{q}-\bar{P};x',x);
\end{equation}
\begin{equation}\label{decay2}
f'_{\pi}\frac{i}{2}\epsilon_{\mu\nu\rho\sigma}eF^{\nu\rho}\bar{P}^{\sigma}=N_ctr_D\int_{x,x',\bar{q}}-i\gamma_\mu S(\bar{q};x,x')\Gamma_{\pi}(\bar{P};x')S(\bar{q}-\bar{P};x',x).
\end{equation}
Since the Schwinger phase factor does not contribute in our simplified vertex of $\Ga_{\pi}$, i.e., two phase factors are involved in the above formulas, we have replaced the derivative of position $\mi\p_{\mu}$ to $\bar{P}_{\mu}$. In our case, for $\bar{P}_{\mu}=(0,0,P_z,0)$ and $F_{12}=-F_{21}=B$, the combination $F_{\mu\nu}\bar{P}_{\nu}$ vanishes automatically. We therefore examine $f$ and  $f'$ in the present paper. Note here that the normalization of BSA is applied by
\begin{equation}\label{norm}
\left[\frac{\partial \ln\lambda(\bar{P}^2)}{\partial \bar{P}^2}\right]^{-1}=-2N_ctr_D\int_{\bar{q},x,x',x''}\bar{\Gamma}_{\pi}(-\bar{P};x-x'')S(\bar{q};x,x')\Gamma_{\pi}(\bar{P};x')S(\bar{q}-\bar{P};x',x).
\end{equation}

\begin{figure}[ht]
\includegraphics[width=0.5\textwidth]{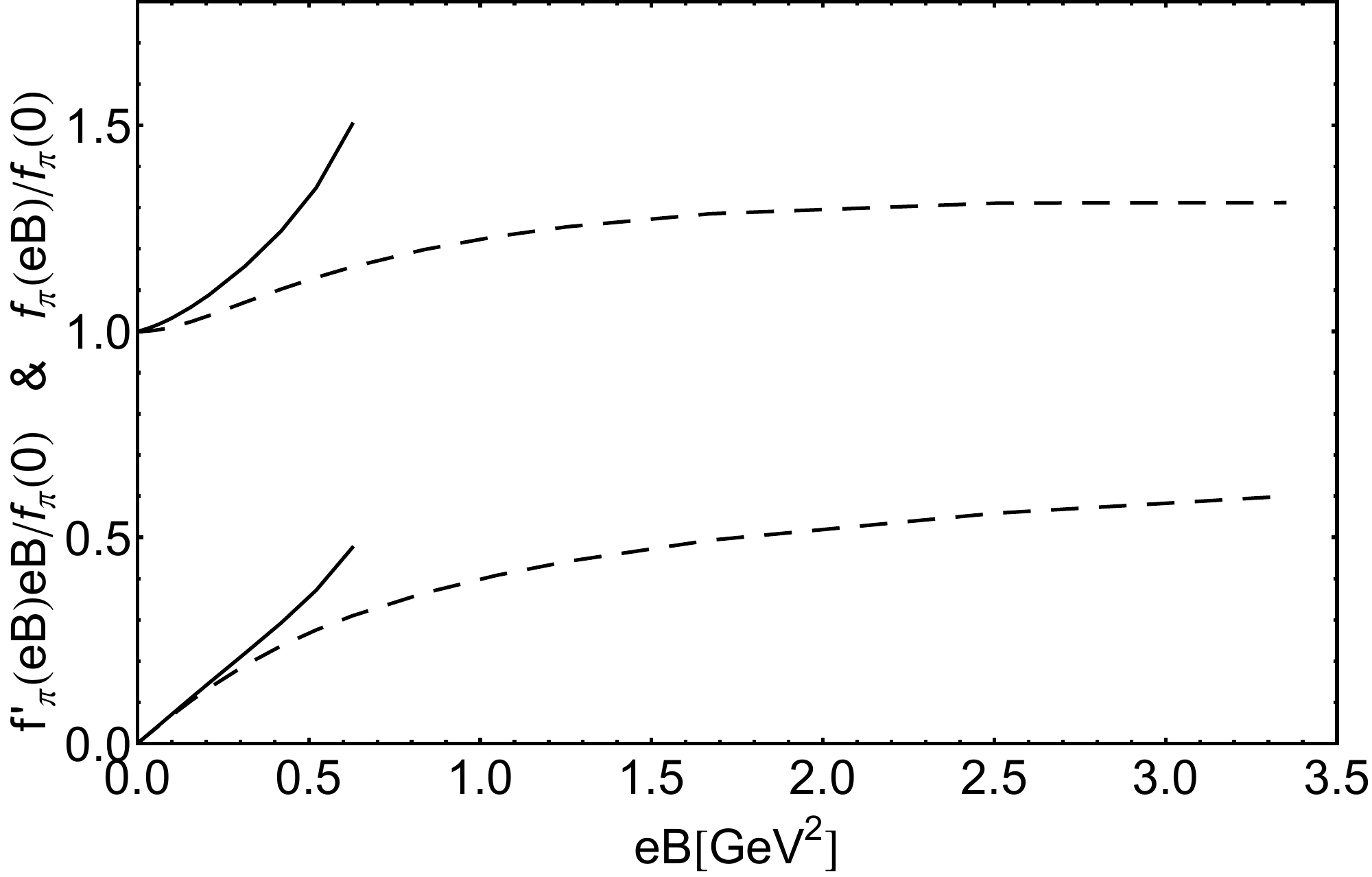}
\caption{Calculated results of the charged pion decay constants as the function of the magnetic field in the conventional contact interaction model. $f(eB)$ are presented in the upper panel and $f'(eB)eB$ are presented in the lower panel.
The solid lines denote the corresponding decay constant computed by a complete BSA incorporating both $E$ and $F$ terms;
the dashed lines represent the ones in terms of a single $E$ component.}\label{charged pion decay result}
\end{figure}

The obtained results of the magnetic field dependence of the charged pion decay constants are shown in~\Fig{charged pion decay result}.
Constructing the BSA in the contexts of both $E$ and $F$, $f$ and $f'$ increases as the magnetic field increases. The trend is slower when a single $E$ component is involved.
%%%%%%%%%%%%%%%%%%%%%%%%%%%%%%%%
%%%%%%%%%%%%%%%%%%%%%%%%%%%%%%%%
\section{When there is an auxiliary term in the contact model}\label{sec:model}
When a single $E$ component of the BSA is applied, the form of the quark-meson vertex is represented by the bare $\gamma_5$ in calculation, which obviously is a bad approximation. It is known that the Bethe-Salpeter wave function is related to the BSA according to
\begin{equation}
\chi_{\pi}(P)=S(q_+)\Gamma(P)S(q_-),
\end{equation}
where $q_+=q+\sigma P$ and $q_-=q-(1-\sigma) P$ with $0<\sigma<1$. From \Eqn{BSA}, the term of $E$ in the BSA contributes as the $s$ wave in the BS wave function while as the axial-vector component of $F$ is attributed to both $s$- and $p$-wave functions, discussed in Refs.~\cite{Hilger:2015ora,Bhagwat:2006xi}. Therefore, within Dyson-Schwinger formalism, this allows us to investigate the internal structures of the meson modes.

Adopting the conventional form of $F$, it has been found that the DSE model calculation fails to present the tight meson as a bound state starting from the moderate magnetic field region. A detailed calculation of pseudoscalar and vector mesons can be found in the thesis work of Ref.~\cite{wang2014}. To address the failure found in that work, we propose to take full account of the backreaction on the gluon sector under the influence of the magnetic field and hence use a more suitable ansatz of the $F$-term in BSA. As explored in Ref.~\cite{Chao:2014wla}, the Lorentz structures of the gluon propagator have been split into three parts in the magnetized vacuum, which takes the form 
\begin{align}
D^{\mu\nu}(k)
=\frac{P_{\sp}(k)}{k^{2}-m^{2}_{G,\sp}}
+\frac{P_{\perp}(k)}{k^{2}-m^{2}_{G,\perp}}
+\frac{k^{\mu}k^{\nu}/k^2-g^{\mu\nu}-P_{\sp}(k)-P_{\perp}(k)}{k^{2}-m^{2}_{G,\mr{mix}}},
\end{align}
where $P_{\sp}(k)=k_{\sp}^{\mu}k_{\sp}^{\nu}/k_{\sp}^2-g^{\mu\nu}_{\sp}$ and $P_{\perp}(k)=k_{\perp}^{\mu}k_{\perp}^{\nu}/k_{\perp}^2-g^{\mu\nu}_{\perp}$. Note here that the metric convention $g^{\mu\nu}$ is decomposed into two orthogonal subspaces $g^{\mu\nu}_{\sp}=\l(1,0,0,-1\r)$ and $g^{\mu\nu}_{\perp}=\l(0,-1,-1,0\r)$. Similar decomposition is adopted for four-dimensional momentum $k_{\mu}=k_{\sp}+k_{\perp}$, $k_{\sp}=\l(k_{0},0,0,k_{3}\r)$, and $k_{\perp}=\l(0,k_{1},k_{2},0\r)$.
It shows that $m^{2}_{G,\sp}\neq m^{2}_{G,\perp}\neq m^{2}_{G,\mr{mix}}$ at finite $eB$. As a consequence, the effect of the $p$-wave in mesons is modified regarding the inclusion of a magnetic field.
To account for the splitting of the gluonic degrees of freedom, we suggest the following replacement of the quark propagator:
\begin{equation}\label{gapeqadd}
\begin{aligned}
S^{-1}\!\left(a,b\right)&=S_0^{-1}\!\left(a,b\right)+\frac{4}{3 m_G^2}\delta^4\!\left(a-b\right)\!\gamma_\mu S(a,b)\gamma_\mu\\
&-\tilde{\eta}^{2}\frac{8}{3 m_G^2}\delta^4\!\left(a-b\right)\!(\gamma_\mu S(a,b)\gamma_\mu+\gamma_5\gamma_\mu S(a,b)\gamma_5\gamma_\mu),
\end{aligned}
\end{equation}
where the explicit form of $\tilde{\eta}$ will be written down in the next context.
Specifically, as an early investigation, we require that the second line in the interaction kernel does not affect the early discussed properties of quarks in the present work.

To maintain Ward-Takahashi identity (WTI), the corresponding BSE is of the form
\begin{equation}\label{BSEadd}
\begin{aligned}
\Gamma\left(\bar{P};x\right)&=-\frac{4}{3 m_G^2}\int_{\bar{q},x'}\!\gamma_\mu S_1\left(\bar{q},x,x'\right)\Gamma\left(\bar{P};x'\right)S_2\left(\bar{k},x',x\right)\gamma_\mu\\
&+\tilde{\eta}^{2}\frac{8}{3 m_G^2}\int_{\bar{q},x'}\!(\gamma_\mu S_1\left(\bar{q},x,x'\right)\Gamma\left(\bar{P};x'\right)S_2\left(\bar{k},x',x\right)\gamma_\mu+\gamma_5\gamma_\mu S_1\left(\bar{q},x,x'\right)\Gamma\left(\bar{P};x'\right)S_2\left(\bar{k},x',x\right)\gamma_5\gamma_\mu).
\end{aligned}
\end{equation}
The auxiliary term controls the strength of the $F$ term by the function of $\tilde{\eta}^{2}$, serving as a tool to exhibit the role of the $p$ wave. To achieve our purpose, the magnitude of $\tilde{\eta}^{2}$ is determined in an empirical way: (i) since the auxiliary term is introduced as the excitation of magnetic fields, it is natural to require $\tilde{\eta}^{2}(eB=0)=0$; (ii) to maintain the spectra of mesons in vacuum, we let $\tilde{\eta}_{0}^{2}=\frac{1}{4}$ to ensure that the $F$ term vanishes at zero magnetic field, which is proportional to $1-4\tilde{\eta}^{2}$ as shown in the following. It also means we assume that the contribution of the $p$ wave is too tiny to remain in vacuum; (iii) the form of $\tilde{\eta}^2$ is equivalent to be brought by a one-meson exchange contact interaction, and it has to be proportional to even powers of $\bar{P}$ to avoid affecting the Goldstone nature in the chiral limit; (iv) after an exploration in the parameter space, we found that, to fully fit the lattice QCD results in Ref.~\cite{Ding2020}, one has
\begin{equation}\label{eta}
\tilde{\eta}^{2}=\tilde{\eta}_{0}^{2}-a*\bar{P}^2*\tau_{uv}^2*\tanh(b* eB^2*\tau_{uv}^4),
\end{equation}
where $a=3.5,\ b=0.153$, and $\tau_{uv}$ is introduced as the energy scale.

Plugging into $\tilde{\eta}^2$, the BSE is expressed as
\begin{equation}\label{neutralpionBSEeta}
\begin{aligned}
\left[\begin{array}{c}E\\F\end{array}\right]=
\frac{4}{3 m_G^2}\left[\begin{array}{cc}1&0\\0& 1-4\tilde{\eta}^{2}\end{array}\right]
\left[\begin{array}{cc}K_{EE}^{\pi_{0}^u}&K_{EF}^{\pi_{0}^u}\\K_{FE}^{\pi_{0}^u}&K_{FF}^{\pi_{0}^u}\end{array}\right]\left[\begin{array}{c}E\\F\end{array}\right].
\end{aligned}
\end{equation}
As noted above, a factor of $(1-4\tilde{\eta}^{2})$ arises in the second line of the kernel matrix. The elements of $K_{EE}^{\pi_{0}^u}, K_{EF}^{\pi_{0}^u}, K_{FE}^{\pi_{0}^u}$ and $K_{FF}^{\pi_{0}^u}$ are exactly the same as those in Eq.~$(\ref{neutralpionBSE})$.

\begin{figure}[ht]
\includegraphics[width=0.5\textwidth]{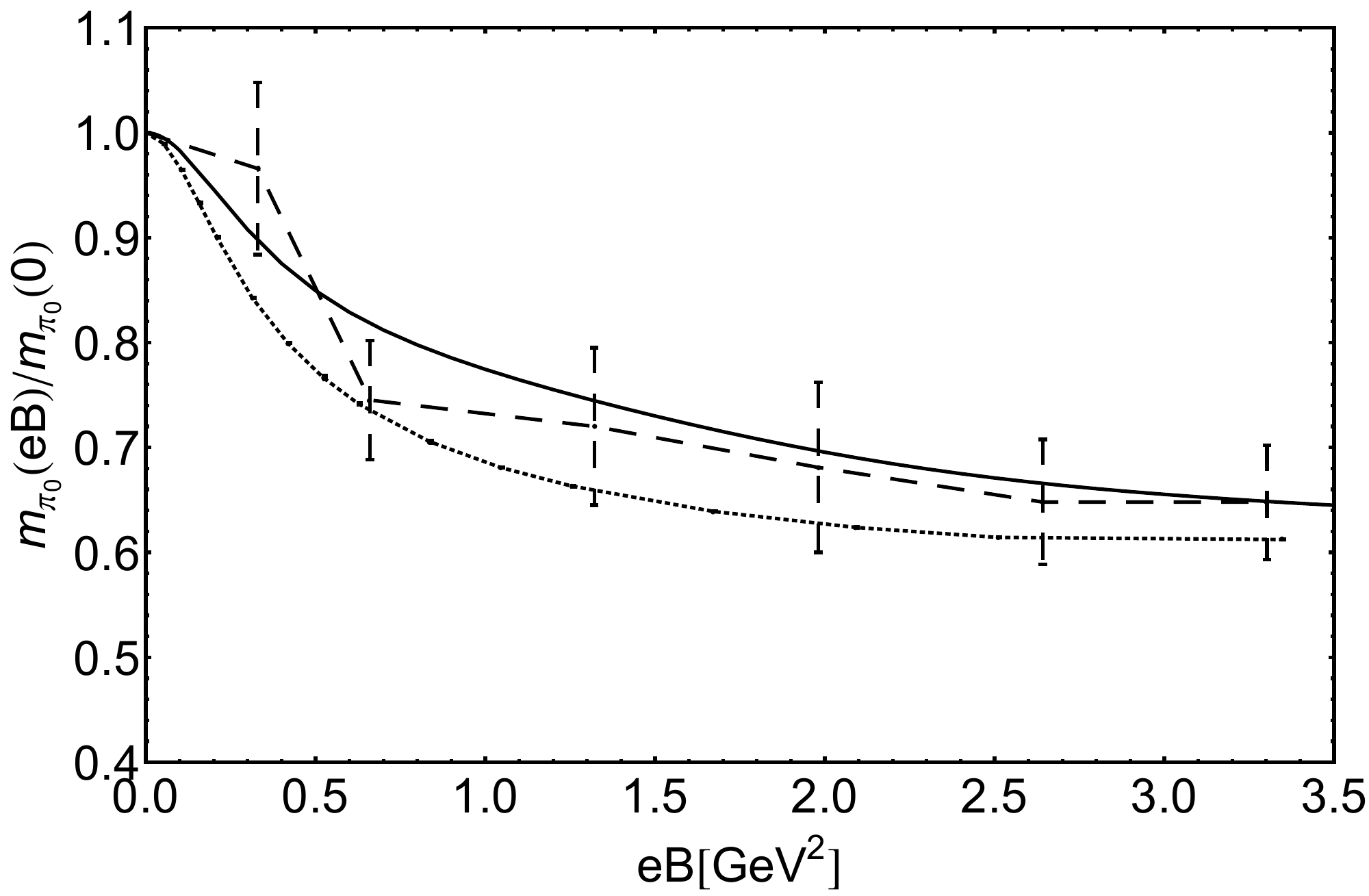}
\caption{Calculated results of neutral pion mass $m_{\pi_{0}^{u}}$ as the function of the magnetic field in the improved contact interaction model. The solid line is computed with $\tilde{\eta}^{2}$ in Eq.~(\ref{eta}); the dashed line is lattice computation from Ref.~\cite{Bali2018}; the dotted line is lattice computation from Ref.~\cite{Ding2020}.}\label{neutral pion mass eta result}
\end{figure}

After solving \Eqn{neutralpionBSEeta}, we have the spectra of neutral pions as a function of the strength of the magnetic field, as shown in~\Fig{neutral pion mass eta result}. It is observed that the mass of $\pi_{0}$ decreases and then tends to saturate, which is roughly in accordance with both the LQCD simulations in Refs.~\cite{Ding2020,Bali2018}. Such trend of $\pi_{0}$ is originated from that $1-4\tilde{\eta}^{2}$ limits to zero as $\bar{P}^2\sim 0$, which diminishes the role of the $F$ component and ceases the incorrect tendency of the mass increasing, which was shown in \Fig{neutral pion mass result}. We note here that, to obtain the results smaller than $eB=0.1~\GeV^2$, we have summed double infinite Landau levels in an analytic manner. The detailed formula are presented in the Appendix~\ref{App:ll}.
%%%%%%%%%%%%%%%%%%%%%%%%%%
%%%%%%%%%%%%%%%%%%%%%%%%%%%%%%%%
\begin{figure}[ht]
\includegraphics[width=0.5\textwidth]{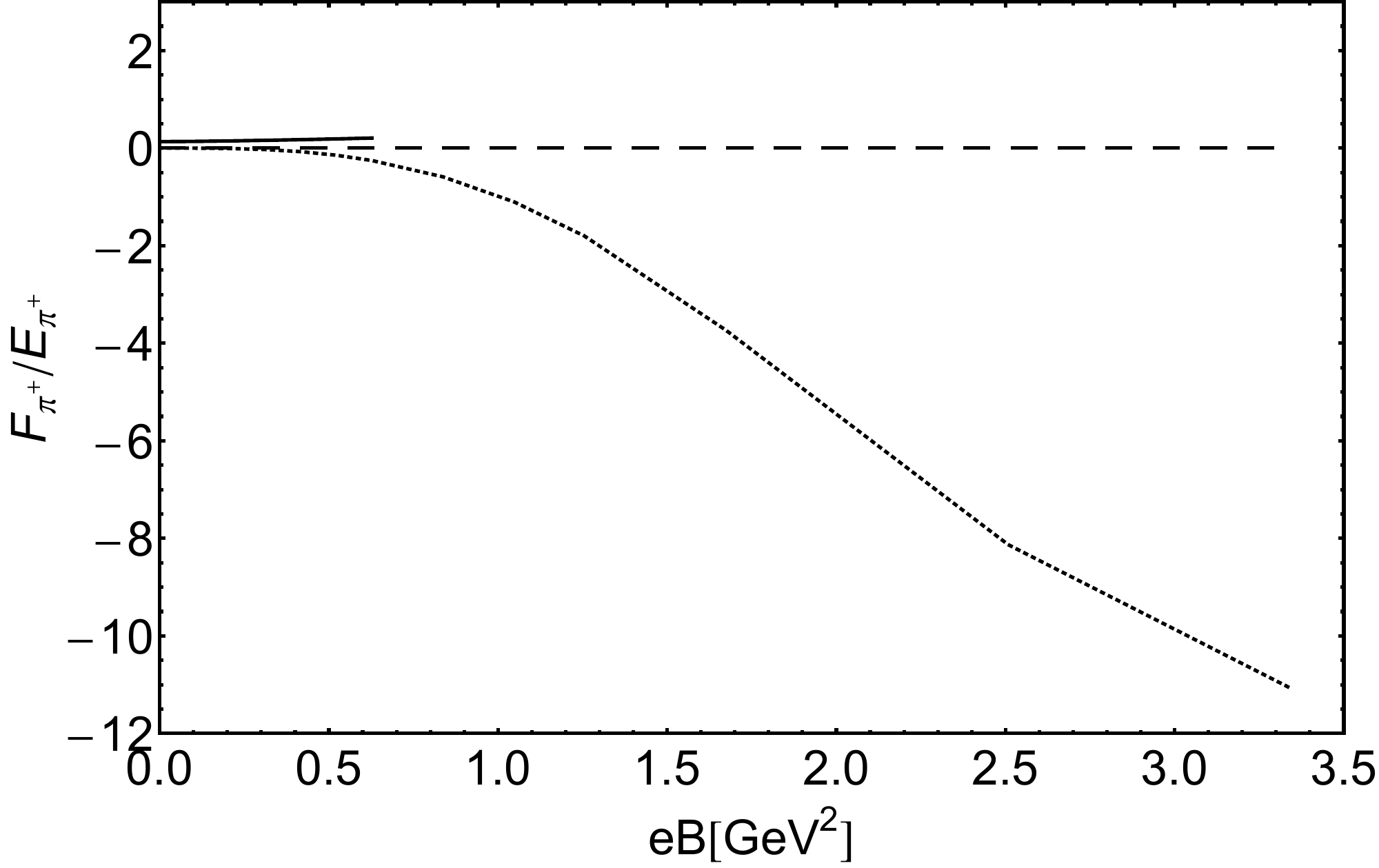}
\caption{Calculated ratio of $E/F$ as the function of the magnetic field with different model parameters. The solid line and dashed line present the results of the full BSA and single $E$, respectively, in the conventional contact model. The dotted line denotes the results in the improved contact interaction model. Here, the general form of the ground state in the BSA is $\sim e^{-|eB|x^2/2}$ for a charged pion, and the line is plotted with the strength ratio between the $E$ and $F$ components.}\label{pion bsa ratio result}
\end{figure}
%%%%%%%%%%%%%%%%%%%%%%%%%%%%%%%%

Similarly, we write down the BSE of charged pions after implementing into $\tilde{\eta}^{2}$,
\begin{equation}\label{chargedpionBSEeta}
\begin{aligned}
\left[\begin{array}{c}E_{\pi_{+}}\\F_{\pi_{+}}\end{array}\right]=\frac{4}{3 m_G^2}\int dx^\prime
\left[\begin{array}{cc}1&0\\0& 1-4\tilde{\eta}^{2}\end{array}\right]
\left[\begin{array}{cc}K_{EE'}^{\pi_{+}}&K_{EF'}^{\pi_{+}}\\
K_{FE'}^{\pi_{+}}& K_{FF'}^{\pi_{+}}\end{array}\right]
\left[\begin{array}{c}E'_{\pi_{+}}\\F'_{\pi_{+}}\end{array}\right].
\end{aligned}
\end{equation}
Here, the kernels $K_{EE'}^{\pi_{+}},\  K_{EF'}^{\pi_{+}},\ K_{FE'}^{\pi_{+}},\ K_{FF'}^{\pi_{+}}$ are exactly the same with those in Eq.~$(\ref{chargedpionBSE})$.
Again, we emphasize that another representation of above kernels is written in the Appendix~\ref{App:ll}, which is used to calculate in the region of $eB<0.1~\GeV^2$.

Under the influence of $\tilde{\eta}^{2}$, one finds out that the relative sign of $E/F$ has been flipped to negative compared with the conventional approach. It is known that the $p$ wave forms as a triplet state, and those three behave the same at finite temperatures or densities. However, the energy states of the charged vector mesons are spin dependent, called the Zeeman effect, and only one of $s_{z}$ reduces the energy dispersion for spin-one meson modes. We numerically confirm that the demanded $p$-wave component is attributed from the negative $F$ term, where the ratio of $E/F$ is shown in~\Fig{pion bsa ratio result}. As discussed in Sec.~\ref{sec:oringal}, the term of $E$ plays a dominant role in the weak limit and the contribution of the $F$ structure is enhanced in moderate and strong regimes of magnetic fields.

\begin{figure}[ht]
\includegraphics[width=0.5\textwidth]{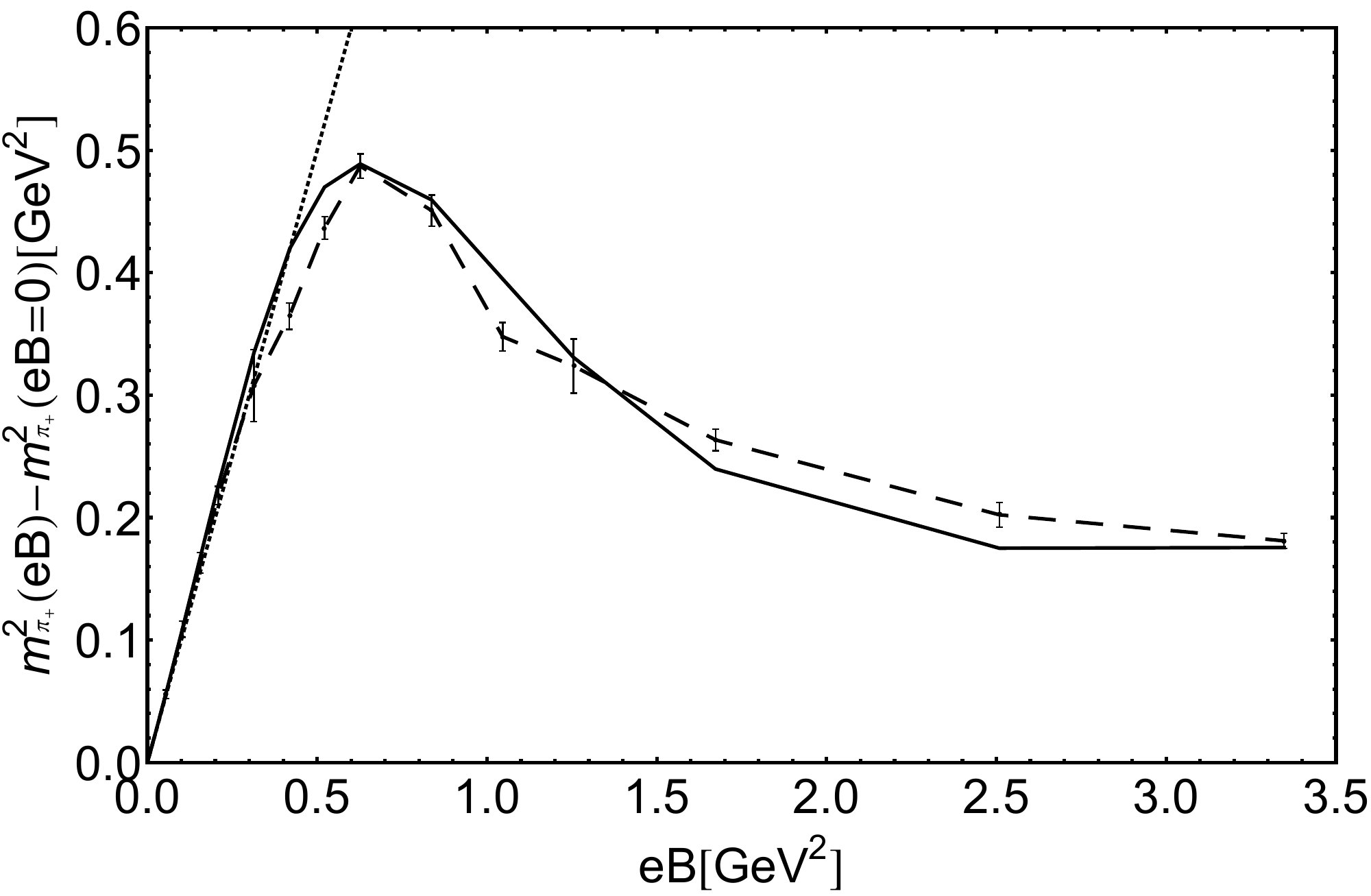}
\caption{Calculated results of charged pion mass as the function of the magnetic field in the improved contact interaction model.
The solid line is computed with $\tilde{\eta}^{2}$ in Eq.~(\ref{eta}), the dotted line denotes the result of LLL approximation, and the dashed line is the lattice result, given in Ref.~\cite{Ding2020}.}\label{charged pion mass eta result}
\end{figure}

\begin{figure}[ht]
\includegraphics[width=0.5\textwidth]{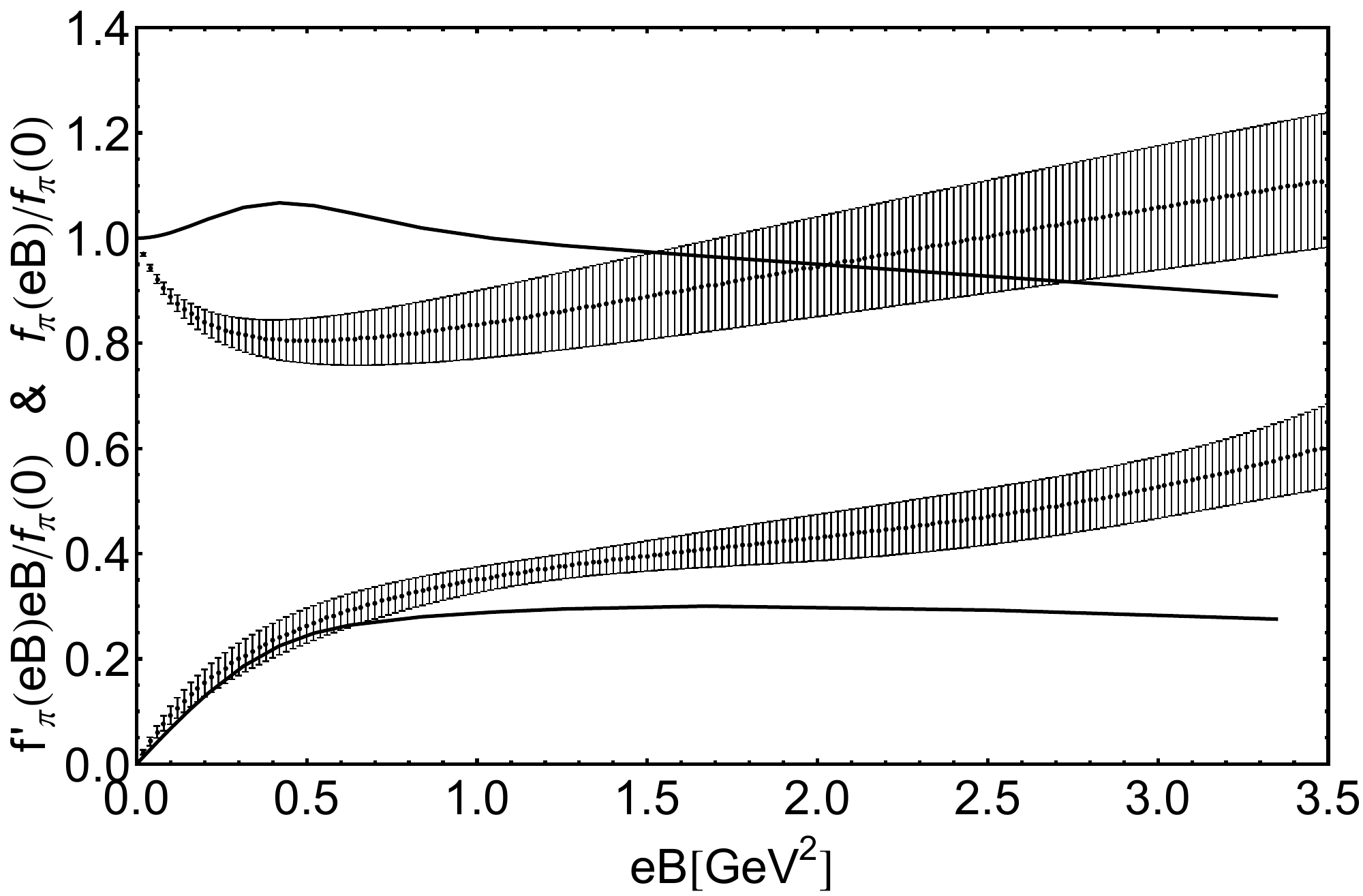}
\caption{Calculated results of the weak decay constant as the function of the magnetic field in the improved contact interaction model.  $f(eB)$ are presented in the upper panel and $f'(eB)eB$ are presented in the lower panel. The solid line is computed with $\tilde{\eta}^{2}$ in Eq.~(\ref{eta}) and the band is the lattice result, given in Ref.~\cite{bali_weak_2018}}
\label{charged pion decay eta result}
\end{figure}

The final results of charged pion masses are plotted in~\Fig{charged pion mass eta result}. Here, we get a nonmonotonic curve as the function of magnetic fields by adopting a nontrivial $\tilde{\eta}^{2}$. One observes that the spectra of the charged pion grows and fits the point-particle description in the weak strength regime of the $eB$ field; the line bends at the greatest point, around $eB\sim 0.6~\GeV^2$ and then decreases as the magnetic field increases, which agrees with the LQCD simulation result in Ref.~\cite{Ding2020}. We also present the weak decay constant $f(eB)$ in~\Fig{charged pion decay eta result}. It shows a similar nonmonotonic behavior like the mass, which is consistent in the present model approach. Another axial decay constant $f'(eB)eB$ saturates while $eB>0.6~\GeV^2$, which is mainly induced by the crude contact interactions. 
We note here that, for $eB<0.1~\GeV^2$, the used formula have been listed in the Appendix~\ref{App:ll}.
A sophisticated investigation on the decay constants is in progress.
%%%%%%%%%%%%%%%%%%%%%%%%%%%%
%%%%%%%%%%%%%%%%%%%%%%%%%%%%
%%%%%%%%%%%%%%%%%%%%%%%%%%%%
\section{Summary}\label{sec:con}
In the present paper, we have systematically studied the pion properties in the magnetized QCD matter under the framework of the Dyson-Schwinger equation. Employing the contact model, we have discussed the spectra of dynamical quarks as well as pion triplets, since these are the fundamental properties of hadron states. Created by the metric $g_{\mu\nu}$ and the additional tensor structures $F_{\mu\nu}$ and $\tilde{F}_{\mu\nu}$, the relevant leptonic decay modes of mesons are three. We have calculated the $f$, relating to $g_{\mu\nu}$ and $f'$, coming with $\tilde{F}_{\mu\nu}$, as the functions of $eB$, while as $f''$, constructing with $F_{\mu\nu}$, vanishes because of the simplified ansatz of $\Gamma_{\pi}$.

In the context of vector-vector contact interaction, $\gamma_{\mu}\otimes\gamma_{\mu}$, the constituent quark mass increases as the magnetic field grows and there is no clue that the charged pion spectra will decrease under the strong magnetic fields. Inspired by the Zeeman splitting in the energy dispersion of charged vector mesons, we suggested that an enhanced $p$-wave state in the axial-scalar meson will reduce the pole mass, making it as light as $\rho^{\pm}$ with spin component $s_z=\mp1$. To demonstrate our conjecture, we have provided a simple method to stimulate the higher angular momentum contribution in the contact model by implementing an auxiliary term, $\gamma_{\mu}\otimes\gamma_{\mu}-\gamma_{\mu}\gamma_{5}\otimes\gamma_{\mu}\gamma_{5}$, into the interactions. In particular, such coupling does not affect the quark sector but is highly involved in the Bethe-Salpeter amplitude of pions.

Before our model computations, for comparisons, we have applied a complete Bethe-Salpeter amplitude to show the quantities of pions without including the auxiliary term. We have obtained that the spectra of $\pi_{0}$ increases with magnetic field, $eB>0.5\,\mathrm{GeV}^2$ while the charged pion mass grows too quickly to achieve a bound state for $eB>0.8\,\mathrm{GeV}^2$. These behaviors indicate that, due to the presence of the magnetic field, the degeneracy of the $p$ wave has been broken and the sign of the $F$ component has become a significant factor to characterize the spin direction. Meanwhile, we have found out that both $f$ and $f'|eB|$ increase with magnetic field for the specified BSA. We have also calculated these quantities by artificially letting $F_{\pi}\left(\bar{P};x\right)=0$ and observed that the BSA reduces to $\Gamma_{\pi}\left(\bar{P};x\right)=\gamma_5E_{\pi}\left(\bar{P};x\right)$, manifesting as a pure $s$-wave state and very close to the calculations of the LLL approximation. The resulting neutral pion mass first decreases and then tends to saturate for $eB>1.5\,\mathrm{GeV}^2$. The energy dispersion of $\pi^{\pm}$ is slightly larger than the point-particle approximation. Moreover, the two remaining decay constants still grow along the magnetic field, but much more slowly than the complete BSA case.

In order to control the strength of the $p$-wave state and fit the LQCD results in Ref.~\cite{Ding2020}, we have specified the value of $\tilde{\eta}^2$. Not surprising, the neutral pion spectra of the present work are in accordance with the first principle simulations results~\cite{Bali2018,Ding2020}. Besides, we have obtained that the mass of charged pions bends around $eB\sim 0.8\,\mathrm{GeV}^2$ and becomes flat in the ultrastrong magnetic fields, which reproduces the curve in Ref.~\cite{Ding2020}. In our model calculations, the charged pion decay constant $f_\pi(eB)/f_\pi(0)$ behaves like the trend of mass and does not deviate from $1$ too much after $eB>0.3\,\mathrm{GeV}^2$ while as  $f'_\pi(eB)|eB|/f_\pi(0)$ increases straightly in the weak limit and then saturates for $eB>0.3\,\mathrm{GeV}^2$. These curves of decay constants do not agree well with the lattice results in Ref.~\cite{bali_weak_2018}. Indeed, at large magnetic field regions, the artificial saturation of the decay constants is inherited from the employed contact interactions. Further research is needed to investigate the strongly magnetized medium.

Apparently, the presently employed model is too simple to manifest the full physics subject to an external magnetic field. The form of $\tilde{\eta}^{2}$ is not derived from an {\it ab initio} approach, but adopted through a phenomenological way. A complete representation of Poincare covariant components in the Bethe-Salpeter equation has not been fulfilled yet. The main aim of the current work, as a first attempt, is to serve as a model to exhibit the effect of the $p$ wave. Therefore, it is still too early to tell the discrepancy between the two nonperturbative computations, while a sophisticated description of the higher angular momentum state under the effect of the magnetic field has not been achieved.

Last but not least, even though the lifetime of magnetic fields in off-central heavy-ions is too short to result in any observations, our analysis of weak decay constants will influence the astrophysical environment and the extension at finite temperatures will be explored in the future.
\\
\hspace{1cm}

\begin{acknowledgements}
We thank H. Ding for sharing their pion masses data and  B.B. Brandt and G. End\"{o}di for sharing their pion masses and weak decay constants data.
This work is supported by the National Natural Science Foundation of China (NSFC)  Grants  No. 11435001, No. 11775041, and No. 12135007.  The work of J. C. has been supported by the start-up funding from Jiangxi Normal University under Grant No. 12021211.
\end{acknowledgements}

\appendix
\section{\label{App:quark}Quark propagator}
In this section, we define $P_\pm = P_\pm^{\xi=1}$ and $I_\pm = I_\pm^{\xi=1}$, and the flavor dependence is suppressed for convenience. The explicit form of the Hermitian operator $G_0^{\prime\,2}\left(a,b\right)$ is
\begin{equation}
\begin{aligned}
G_0^{\prime\,2}\left(a,b\right)=\delta^4(a-b)\left[-\partial_b^2-2ie_fA\cdot\partial-ie_f(\partial_\mu A_\nu)\gamma_\mu\gamma_\nu+e_f^2A^2+M^2\right].
\end{aligned}
\end{equation}
By solving Eq.~(\ref{eigeneq}), we can obtain the eigenfunctions in $(x,p_y,p_z,p_t)$ space
\begin{equation}\label{AEq:eigenfun}
\begin{aligned}
P_{\pm}f_{n,\bar{p}}\left(b\right)=P_{\pm}\hat{f_n}\left(b_x+\frac{p_y}{eB}\right)e^{\left(i\bar{b}\cdot\bar{p}\right)},
\end{aligned}
\end{equation}
where $\bar{p}=(p_y,p_z,p_t),\bar{b}=(b_y,b_z,b_t)$. The Fourier factor $e^{\left(i\bar{b}\cdot\bar{p}\right)}$ indicates the translation invariant $(y, z, t)$ directions and $\hat{f}_n\left(x\right)$ is the normalized harmonic oscillator wave function in Eq.~(\ref{nhwf}). It is convenient to introduce ladder operators
\begin{equation}
\begin{aligned}
\hat{a}_+=\frac{\left|eB\right|x-\partial_x}{\sqrt{2\left|eB\right|}},\quad\hat{a}_-=\frac{\left|eB\right|x+\partial_x}{\sqrt{2\left|eB\right|}},
\end{aligned}
\end{equation}
where
\begin{equation}
\begin{aligned}
\hat{a}_+\hat{f}_n\left(x\right)=\sqrt{n+1}\hat{f}_{n+1}\left(x\right),\quad\hat{a}_-\hat{f}_n\left(x\right)=\sqrt{n}\hat{f}_{n-1}\left(x\right).
\end{aligned}
\end{equation}
Forming as a complete orthogonal basis, we use the eigenfunctions (\ref{AEq:eigenfun}) to expand the inverse of the dressed propagator (\ref{inversedressedpropagator}). The expansion coefficients are
\begin{equation}
\int d^4b S^{-1}\left(a,b\right)P_{\pm}f_n\left(b\right)=\left(i\slashed{p}+M+I_+\left(\partial_{\hat{x}}+eB\hat{x}\right)+I_-\left(\partial_{\hat{x}}-eB\hat{x}\right)\right)P_{\pm}f_n\left(a\right),
\end{equation}
where ${\hat{x}}=a_x+\frac{p_y}{eB}$, $p=(0,0,p_z,p_t)$, $I_\pm = I_\pm^{\xi=1}$, and $I_{\pm}^{\xi}=\frac{\gamma_1\pm i\xi\gamma_2}{2}$. $f_{n,\bar{p}}\left(a\right)$ is written as $f_n\left(a\right)$ for convenience. For $eB>0$:
\begin{equation}
\int db S^{-1}\left(a,b\right)P_{+}f_{n}\left(b\right)=P_{+}\left(i\slashed{p}+M\right)f_{n}\left(a\right)+I_+\sqrt{2B_{n}}f_{n-1}\left(a\right),
\end{equation}
\begin{equation}
\int d b S^{-1}\left(a,b\right)P_-f_n\left(b\right)=P_-\left(i\slashed{p}+M\right)f_n\left(a\right)-I_-\sqrt{2B_{n+1}}f_{n+1}\left(a\right).
\end{equation}
Using $\int d^4b d^4c S^{-1}\left(a,b\right) S\left(b,c\right)P_\pm f_{n}\left(c\right)=P_\pm f_{n}\left(a\right)$, the expansion coefficients of the dressed propagator are
\begin{equation}
\begin{aligned}
\int d b S\left(a,b\right)&P_+f_n\left(b\right)=\left(P_+\left(-i\slashed{p}+M\right)f_n\left(a\right)-I_+\sqrt{2B_{n}}f_{n-1}\left(a\right)\right)\frac{1}{p^2+M^2+2B_{n}},
\end{aligned}
\end{equation}
\begin{equation}
\begin{aligned}
\int d b S\left(a,b\right)&P_-f_n\left(b\right)=\left(P_-\left(-i\slashed{p}+M\right)f_n\left(a\right)+I_-\sqrt{2B_{n+1}}f_{n+1}\left(a\right)\right)\frac{1}{p^2+M^2+2B_{n+1}}.
\end{aligned}
\end{equation}
Similarly, we obtain the expansion coefficients of the propagator for $eB<0$. We summarize the propagator with any $eB$ as
\begin{equation}
\begin{aligned}
S\left(\bar{p};x,x^\prime\right)=&\sum_{n=0}^{\infty}{\left(P_{+}^{\xi}\left(-i\slashed{p}+M\right)f_n\left(x,p_y\right)-I_{+}^{\xi}\sqrt{2B_{n}}f_{n-1}\left(x,p_y\right)\right)\frac{1}{p^2+M^2+2B_{n}}}P_{+}^{\xi}f_n\left(x^\prime,p_y\right)\\
+&\left(P_{-}^{\xi}\left(-i\slashed{p}+M\right)f_n\left(x,p_y\right)+I_{-}^{\xi}\sqrt{2B_{n+1}}f_{n+1}\left(x,p_y\right)\right)\frac{1}{p^2+M^2+2B_{n+1}} P_{-}^{\xi}f_n\left(x^\prime,p_y\right).
\end{aligned}
\end{equation}
Now $|eB|$ appears everywhere except in the harmonic wave function $f_n\left(x,p_y\right)=\hat{f}_n\left(x+\frac{p_y}{eB}\right)$. Finally, for the dynamical quark mass
\begin{equation}
\begin{aligned}
M&=m_0+\frac{4}{3 m_G^2}\gamma_\mu S(a,a)\gamma_\mu,
\end{aligned}
\end{equation}
the propagator $S(a,a)$ is explicitly expressed as
\begin{equation}
\begin{aligned}
S\left(a,a\right)&=\int_{\bar{p}}\sum_{n=0}^{\infty}{\left(\frac{P_{+}^{\xi}\left(-i\slashed{p}+M\right)}{p^2+M^2+2B_{n}}f_n^{2}\left(a_x,p_y\right)+\frac{P_{-}^{\xi}\left(-i\slashed{p}+M\right)}{p^2+M^2+2B_{n+1}}f_n^{2}\left(a_x,p_y\right)\right)}e^{i\bar{p}\cdot(\bar{a}-\bar{a})}\\
&=\int_{p}\sum_{n=0}^{\infty}{\frac{\left|eB\right|}{2\pi}\left(\frac{P_{+}^{\xi}\left(-i\slashed{p}+M\right)}{p^2+M^2+2B_{n}}+\frac{P_{-}^{\xi}\left(-i\slashed{p}+M\right)}{p^2+M^2+2B_{n+1}}\right)}.
\end{aligned}
\end{equation}

\section{\label{App:BSE}Meson properties}
\subsection{Neutral pion}
The kernels in Eq.~(\ref{neutralpionBSE}) are formed as
\begin{equation}
\begin{aligned}
K_{EE}^{\pi_{0}^u}
&=-tr\int dx\int\frac{d^3\bar{q}}{\left(2\pi\right)^3}{\frac{\gamma_5}{4}\gamma_\mu S_u\left(\bar{q},x,0\right)\gamma_5S_u\left(\bar{q}-\bar{P},0,x\right)\gamma_\mu},
\end{aligned}
\end{equation}
\begin{equation}
\begin{aligned}
K_{EF}^{\pi_{0}^u}
&=-tr\int dx\int\frac{d^3\bar{q}}{\left(2\pi\right)^3}{\frac{\gamma_5}{4}\gamma_\mu S_u\left(\bar{q},x,0\right)\frac{-i\gamma_5\slashed{\bar{P}}}{M_u}S_u\left(\bar{q}-\bar{P},0,x\right)\gamma_\mu},
\end{aligned}
\end{equation}\begin{equation}
\begin{aligned}
K_{FE}^{\pi_{0}^u}
&=-tr\int dx\int\frac{d^3\bar{q}}{\left(2\pi\right)^3}{\frac{-i\gamma_5\slashed{\bar{P}}M_u}{4\bar{P}^2}\gamma_\mu S_u\left(\bar{q},x,0\right)\gamma_5S_u\left(\bar{q}-\bar{P},0,x\right)\gamma_\mu},
\end{aligned}
\end{equation}\begin{equation}
\begin{aligned}
K_{FF}^{\pi_{0}^u}
&=-tr\int dx\int\frac{d^3\bar{q}}{\left(2\pi\right)^3}{\frac{-i\gamma_5\slashed{\bar{P}}M_u}{4\bar{P}^2}\gamma_\mu S_u\left(\bar{q},x,0\right)\frac{-i\gamma_5\slashed{\bar{P}}}{M_u}S_u\left(\bar{q}-\bar{P},0,x\right)\gamma_\mu}.
\end{aligned}
\end{equation}
For neutral pion, the integration over $q_y$ and $x$ can be done analytically by the orthogonality of the normalized harmonic oscillator wave function
\begin{equation}
\begin{aligned}
&g_{1,n_1,n_2}=\int dx\int\frac{dq_y}{2\pi}\hat{f}_{n_1}\left(x+\frac{q_y}{e_uB}\right)\hat{f}_{n_1}\left(\frac{q_y}{e_uB}\right)\hat{f}_{n_2}\left(\frac{q_y}{e_uB}\right)\hat{f}_{n_2}\left(x+\frac{q_y}{e_uB}\right)=\frac{\left|e_uB\right|}{2\pi}\delta_{n_1,n_2},\\
&g_{2,n_1,n_2}=\int dx\int\frac{dq_y}{2\pi}\hat{f}_{n_1+1}\left(x+\frac{q_y}{e_uB}\right)\hat{f}_{n_1}\left(\frac{q_y}{e_uB}\right)\hat{f}_{n_2-1}\left(\frac{q_y}{e_uB}\right)\hat{f}_{n_2}\left(x+\frac{q_y}{e_uB}\right)=\frac{\left|e_uB\right|}{2\pi}\delta_{n_1+1,n_2},\\
&g_{3,n_1,n_2}=\int dx\int\frac{dq_y}{2\pi}\hat{f}_{n_1-1}\left(x+\frac{q_y}{e_uB}\right)\hat{f}_{n_1}\left(\frac{q_y}{e_uB}\right)\hat{f}_{n_2+1}\left(\frac{q_y}{e_uB}\right)\hat{f}_{n_2}\left(x+\frac{q_y}{e_uB}\right)=\frac{\left|e_uB\right|}{2\pi}\delta_{n_1-1,n_2},\\
&g_{4,n_1,n_2}=\int dx\int\frac{dq_y}{2\pi}\hat{f}_{n_1}\left(x+\frac{q_y}{e_uB}\right)\hat{f}_{n_1}\left(\frac{q_y}{e_uB}\right)\hat{f}_{n_2}\left(\frac{q_y}{e_uB}\right)\hat{f}_{n_2}\left(x+\frac{q_y}{e_uB}\right)=\frac{\left|e_uB\right|}{2\pi}\delta_{n_1,n_2},
\end{aligned}
\end{equation}
thus, these kernels can be written as
\begin{equation}\label{p0kernel}
\begin{aligned}
K_{EE}^{\pi_{0}^u}&=2\sum^{\infty}_{n_1=0}\sum^{\infty}_{n_2=0}G_{1,n_1,n_2}g_{1,n_1,n_2}+G_{2,n_1,n_2}g_{2,n_1,n_2}+G_{3,n_1,n_2}g_{3,n_1,n_2}+G_{4,n_1,n_2}g_{4,n_1,n_2}\\
&=\frac{\left|e_uB\right|}{2\pi}2\sum^{\infty}_{n_u=0}G_{1,n_u,n_u}+G_{2,n_u,n_u+1}+G_{3,n_u,n_u-1}+G_{4,n_u,n_u}\\
&\equiv\frac{\left|e_uB\right|}{2\pi}2\sum^{\infty}_{n_u=0}G_{1,n_u}+G_{2,n_u}+G_{3,n_u}+G_{4,n_u},\\
K_{EF}^{\pi_{0}^u}&=\frac{\left|e_uB\right|}{2\pi}2\sum^{\infty}_{n=0}G_{5,n_u}+G_{6,n_u},\\
K_{FE}^{\pi_{0}^u}&=\frac{M_{u}^2}{2\bar{P}^2}K_{EF}^{\pi_{0}^{u}},\\
K_{FF}^{\pi_{0}^u}&=-2K_{FE}^{\pi_{0}^{u}}.
\end{aligned}
\end{equation}
In general, the form of $K_{FF}^{\pi_{+}}$ is deduced from WTIs. However, the WTIs in a magnetic field are very complicated; we use the subtraction ($K_{FF}^{\pi_{0}^u}=-2K_{FE}^{\pi_{0}^{u}}$), found in the zero magnetic field, as an approximation~\cite{GutierrezGuerrero2010}. $G_{j,n}$ corresponds to the integration over $q_z, q_t$ and should be regularized by Eq.~(\ref{reg1}), where
\begin{equation}
\begin{aligned}
&G_{1,n_u}=\int\frac{d^2 q}{\left(2\pi\right)^2}\frac{q\cdot k+M_u^2}{(q^2+M_{n_u}^2)(k^2+M_{n_u}^2)}\\
&G_{2,n_u}=\int\frac{d^2 q}{\left(2\pi\right)^2}\frac{2\sqrt{B_{n_u+1}B_{n_u+1}}}{(q^2+M_{n_u+1}^2)(k^2+M_{n_u+1}^2)}\\
&G_{3,n_u}=\int\frac{d^2 q}{\left(2\pi\right)^2}\frac{2\sqrt{B_{n_u}B_{n_u}}}{(q^2+M_{n_u}^2)(k^2+M_{n_u}^2)}\\
&G_{4,n_u}=\int\frac{d^2 q}{\left(2\pi\right)^2}\frac{q\cdot k+M_u^2}{(q^2+M_{n_u+1}^2)(k^2+M_{n_u+1}^2)}\\
&G_{5,n_u}=\int\frac{d^2 q}{\left(2\pi\right)^2}\frac{q\cdot P-k\cdot P}{(q^2+M_{n_u+1}^2)(k^2+M_{n_u+1}^2)}\\
&G_{6,n_u}=\int\frac{d^2 q}{\left(2\pi\right)^2}\frac{q\cdot P-k\cdot P}{(q^2+M_{n_u}^2)(k^2+M_{n_u}^2)}.
\end{aligned}
\end{equation}

\subsection{Charged pion}
The applied kernels in Eq.~(\ref{chargedpionBSE}) read as
\begin{equation}
\begin{aligned}
K_{EE'}^{\pi_{+}}
&=-tr\int\frac{d^3\bar{q}}{\left(2\pi\right)^3}{\frac{\gamma_5}{4}\gamma_\mu S_u\left(\bar{q},x,x^\prime\right)\gamma_5S_d\left(\bar{q}-\bar{P},x^\prime,x\right)\gamma_\mu},
\end{aligned}
\end{equation}
\begin{equation}
\begin{aligned}
K_{EF'}^{\pi_{+}}
&=-tr\int\frac{d^3\bar{q}}{\left(2\pi\right)^3}{\frac{\gamma_5}{4}\gamma_\mu S_u\left(\bar{q},x,x^\prime\right)\frac{-i\gamma_5\slashed{\bar{P}}}{2M}S_d\left(\bar{q}-\bar{P},x^\prime,x\right)\gamma_\mu},
\end{aligned}
\end{equation}\begin{equation}
\begin{aligned}
K_{FE'}^{\pi_{+}}
&=-tr\int\frac{d^3\bar{q}}{\left(2\pi\right)^3}{\frac{-i\gamma_5\slashed{\bar{P}}2M}{4\bar{P}^2}\gamma_\mu S_u\left(\bar{q},x,x^\prime\right)\gamma_5S_d\left(\bar{q}-\bar{P},x^\prime,x\right)\gamma_\mu},
\end{aligned}
\end{equation}\begin{equation}
\begin{aligned}
K_{FF'}^{\pi_{+}}
&=-tr\int\frac{d^3\bar{q}}{\left(2\pi\right)^3}{\frac{-i\gamma_5\slashed{\bar{P}}2M}{4\bar{P}^2}\gamma_\mu S_u\left(\bar{q},x,x^\prime\right)\frac{-i\gamma_5\slashed{\bar{P}}}{2M}S_d\left(\bar{q}-\bar{P},x^\prime,x\right)\gamma_\mu},
\end{aligned}
\end{equation}
and can be rewritten as
\begin{equation}\label{eqn:kernel}
\begin{aligned}
K_{EE'}^{\pi_{+}}&=2\sum^{\infty}_{n_u=0}\sum^{\infty}_{n_d=0}G_{1,n_u,n_d}g_{1,n_u,n_d}+G_{2,n_u,n_d}g_{2,n_u,n_d}+G_{3,n_u,n_d}g_{3,n_u,n_d}+G_{4,n_u,n_d}g_{4,n_u,n_d},\\
K_{EF'}^{\pi_{+}}&=2\sum^{\infty}_{n_u=0}\sum^{\infty}_{n_d=0}G_{5,n_u,n_d}g_{1,n_u,n_d}+G_{6,n_u,n_d}g_{4,n_u,n_d},\\
K_{FE'}^{\pi_{+}}&=\frac{2M^2}{\bar{P^2}}K_{EF'}^{\pi_{+}},\\
K_{FF'}^{\pi_{+}}&=-\frac{M_u+M_d}{2M}K_{FE'}^{\pi_{+}}
\end{aligned}
\end{equation}
where the form of $K_{FF'}^{\pi_{+}}$ is because of the subtraction from WTIs.
$g_{j,n_u,n_d}$ correspond to the integration over $q_y$ and carry the information of $x,x'$, where
\begin{equation}
\begin{aligned}
&g_{1,n_u,n_d}=\int\frac{dq_y}{2\pi}\hat{f}_{n_u}\left(x+\frac{q_y}{e_uB}\right)\hat{f}_{n_u}\left(x'+\frac{q_y}{e_uB}\right)\hat{f}_{n_d}\left(x'+\frac{q_y}{e_dB}\right)\hat{f}_{n_d}\left(x+\frac{q_y}{e_dB}\right),\\
&g_{2,n_u,n_d}=\int\frac{dq_y}{2\pi}\hat{f}_{n_u+1}\left(x+\frac{q_y}{e_uB}\right)\hat{f}_{n_u}\left(x'+\frac{q_y}{e_uB}\right)\hat{f}_{n_d+1}\left(x'+\frac{q_y}{e_dB}\right)\hat{f}_{n_d}\left(x+\frac{q_y}{e_dB}\right),\\
&g_{3,n_u,n_d}=\int\frac{dq_y}{2\pi}\hat{f}_{n_u-1}\left(x+\frac{q_y}{e_uB}\right)\hat{f}_{n_u}\left(x'+\frac{q_y}{e_uB}\right)\hat{f}_{n_d-1}\left(x'+\frac{q_y}{e_dB}\right)\hat{f}_{n_d}\left(x+\frac{q_y}{e_dB}\right),\\
&g_{4,n_u,n_d}=\int\frac{dq_y}{2\pi}\hat{f}_{n_u}\left(x+\frac{q_y}{e_uB}\right)\hat{f}_{n_u}\left(x'+\frac{q_y}{e_uB}\right)\hat{f}_{n_d}\left(x'+\frac{q_y}{e_dB}\right)\hat{f}_{n_d}\left(x+\frac{q_y}{e_dB}\right).
\end{aligned}
\end{equation}
$G_{j,n_u,n_d}$ correspond to the integration over $q_z, q_t$ and should be regularized through Eq.~$(\ref{reg1})$, where
\begin{equation}\label{chargedG}
\begin{aligned}
&G_{1,n_u,n_d}=\int\frac{d^2 q}{\left(2\pi\right)^2}\frac{q\cdot k+M_u M_d}{(q^2+M_{n_u}^2)(k^2+M_{n_d+1}^2)}\\
&G_{2,n_u,n_d}=\int\frac{d^2 q}{\left(2\pi\right)^2}\frac{-2\sqrt{B_{n_u+1}B_{n_d+1}}}{(q^2+M_{n_u+1}^2)(k^2+M_{n_d+1}^2)}\\
&G_{3,n_u,n_d}=\int\frac{d^2 q}{\left(2\pi\right)^2}\frac{-2\sqrt{B_{n_u}B_{n_d}}}{(q^2+M_{n_u}^2)(k^2+M_{n_d}^2)}\\
&G_{4,n_u,n_d}=\int\frac{d^2 q}{\left(2\pi\right)^2}\frac{q\cdot k+M_u M_d}{(q^2+M_{n_u+1}^2)(k^2+M_{n_d}^2)}\\
&G_{5,n_u,n_d}=\frac{1}{2M}\int\frac{d^2 q}{\left(2\pi\right)^2}\frac{q\cdot P M_d-k\cdot P M_u}{(q^2+M_{n_u+1}^2)(k^2+M_{n_d}^2)}\\
&G_{6,n_u,n_d}=\frac{1}{2M}\int\frac{d^2 q}{\left(2\pi\right)^2}\frac{q\cdot P M_d-k\cdot P M_u}{(q^2+M_{n_u}^2)(k^2+M_{n_d+1}^2)}.
\end{aligned}
\end{equation}

We also give the detailed charged pion decay constants after including the auxiliary term.
Firstly, the normalization in Eq.~(\ref{norm}) can be completed as
\begin{eqnarray}
\left[\frac{\partial \ln(\lambda(\bar{P}^2))}{\partial \bar{P}^2}\right]^{-1}&=&-2N_c\int_{x,x',x''}\sum_{X=E,F}\left(K_{EX'}^{\pi_{+}}X(x')E(x-x'')+\frac{1}{2M}\frac{\bar{P}^2}{M}K_{FX'}^{\pi_{+}}X(x')F(x-x'')\right)\nonumber\\
&=&-2N_c\int_{x,x''}\frac{3m_G^2}{4}E(x)E(x-x'')+\frac{3m_G^2}{4}\frac{\bar{P}^2}{2(1-4\tilde{\eta}^2)M^2}F(x)F(x-x'')\nonumber\\
&=&-2N_c\frac{3m_G^2}{4}\frac{2 \pi}{|eB|}(a_E^2+\frac{\bar{P}^2}{2(1-4\tilde{\eta}^2)M^2}a_F^2).
\end{eqnarray}
The ground states of the charged pion take the form of $X(x)=a_Xe^{-|eB|x^2/2}$ for $X=E,F$ to complete the integration of the last second line. Then, the decay constants $f$ and $f'$ are given by
\begin{eqnarray}
f_{\pi}&&=\frac{N_c}{\bar{P}^2}\int_{x,x'}\frac{\bar{P}^2}{M}\left[K_{FE'}^{\pi_{+}}E(x')+K_{FF'}^{\pi_{+}}F(x')\right]\nonumber\\
&&=\frac{3m_G^2}{4}\frac{N_c}{(1-4\tilde{\eta}^2)M}\sqrt{\frac{2\pi}{|eB|}}a_F,
\end{eqnarray}
\begin{equation}
f'_{\pi}=\frac{N_c}{\left|eB\right|\bar{P}^2}\int_{x,x'}\frac{\bar{P}^2}{M}\left[\tilde{K}_{FE'}^{\pi_{+}}E(x')+\tilde{K}_{FF'}^{\pi_{+}}F(x')\right],
\end{equation}
where
\begin{eqnarray}
\tilde{K}_{FE'}^{\pi_{+}}&&=\frac{4M^2}{\bar{P^2}}\sum^{\infty}_{n_u=0}\sum^{\infty}_{n_d=0}G_{5,n_u,n_d}g_{1,n_u,n_d}-G_{6,n_u,n_d}g_{4,n_u,n_d},\\
\tilde{K}_{FF'}^{\pi_{+}}&&=-\frac{M_u+M_d}{2M}\tilde{K}_{FE'}^{\pi_{+}}.
\end{eqnarray}
By analogy with $K_{FF'}^{\pi_{+}}$, the form of $\tilde{K}_{FF'}^{\pi_{+}}$ also comes from the subtraction of WTIs.

\section{\label{App:ll}Summing over the infinite Landau levels}
In the region of strong magnetic fields with~$eB\geq 0.1\,\GeV^2$, we truncate the Landau level at $n=50$. However, such finite summation of Landau levels is far from enough at the weaker region of magnetic fields. To complete the calculations in any magnetic fields, we present below the explicit expressions for the quark gap equation and pion kernels after summing over all Landau levels, which also make it possible to freely choose the scheme of regularization~\cite{Avancini:2015ady,Avancini:2019wed}.
\subsection{Quark and neutral pion}
In the representation of the proper time method, the summation over all Landau levels is achieved by the hyperbolic functions. The quark gap equation is presented as
\begin{equation}\label{eqn:summedgap}
M_f=m_0+\frac{M_f\left|e_fB\right|}{3 \pi^2 m_G^2}\int_{\tau_{uv}^2}^{\tau_{ir}^2} d\tau \frac{e^{-\tau M^2}}{\tau}\coth(\left|e_fB\right| \tau).
\end{equation}
The BSA kernels of neutral pions are rewritten to
\begin{equation}\label{eqn:summedneutral}
\begin{aligned}
K_{EE}^{\pi_{0}^u}&=\frac{\left|e_uB\right|}{4 \pi^2}\int_{\tau_{uv}^2}^{\tau_{ir}^2} d\tau\int_{0}^{1}d\alpha\ \frac{e^{-\tau(M_u^2+\alpha\hat{\alpha}\bar{P}^2)}(1-2\alpha\hat{\alpha}\bar{P}^2\tau)}{\tau}\coth(\left|e_uB\right|\tau),\\
K_{EF}^{\pi_{0}^u}&=\frac{\left|e_uB\right|}{4 \pi^2}\int_{\tau_{uv}^2}^{\tau_{ir}^2} d\tau\int_{0}^{1}d\alpha\ e^{-\tau(M_u^2+\alpha\hat{\alpha}\bar{P}^2)}\bar{P}^2\coth(\left|e_uB\right|\tau),\\
K_{FE}^{\pi_{0}^u}&=\frac{M_{u}^2}{2\bar{P}^2}K_{EF}^{\pi_{0}^{u}},\\
K_{FF}^{\pi_{0}^u}&=-2K_{FE}^{\pi_{0}^{u}},
\end{aligned}
\end{equation}
where $\hat{\alpha}=1-\alpha$.

\subsection{Charged pion}
The ground state of charged pions is modified by the magnetic field. After observation, we assume its BSA is in the form of $\Gamma_{\pi}\left(\bar{P};x\right)=\sqrt{\frac{|eB|}{2\pi}}e^{-|eB|x^2/2}\Gamma_{\pi}\left(\bar{P}\right)$, where $\sqrt{\frac{|eB|}{2\pi}}$ is the normalization factor and $\int_{-\infty}^{\infty}dx\ \Gamma_{\pi}\left(\bar{P};x\right)=\Gamma_{\pi}\left(\bar{P}\right)$. Under the help of the Ansatz, it allows us to integrate over $x$, $x'$, and $q_y$ analytically, the same as the neutral pion. Plus, given by the structures of BSA kernels in~\eqn{eqn:kernel}, one notices that the forms of $G_{j,n_u,n_d}$ are unchanged, seen in~\eqn{chargedG}. The modifications are carried by $g_{j,n_u,n_d}$, where
\begin{equation}
\begin{aligned}
&g_{1,n_u,n_d}=\int dx \int dx'\int\frac{dq_y}{2\pi}\sqrt{\frac{|eB|}{2\pi}}e^{-|eB|x'^2/2}\hat{f}_{n_u}\left(x+\frac{q_y}{e_uB}\right)\hat{f}_{n_u}\left(x'+\frac{q_y}{e_uB}\right)\hat{f}_{n_d}\left(x'+\frac{q_y}{e_dB}\right)\hat{f}_{n_d}\left(x+\frac{q_y}{e_dB}\right),\\
&g_{2,n_u,n_d}=\int dx \int dx'\int\frac{dq_y}{2\pi}\sqrt{\frac{|eB|}{2\pi}}e^{-|eB|x'^2/2}\hat{f}_{n_u+1}\left(x+\frac{q_y}{e_uB}\right)\hat{f}_{n_u}\left(x'+\frac{q_y}{e_uB}\right)\hat{f}_{n_d+1}\left(x'+\frac{q_y}{e_dB}\right)\hat{f}_{n_d}\left(x+\frac{q_y}{e_dB}\right),\\
&g_{3,n_u,n_d}=\int dx \int dx'\int\frac{dq_y}{2\pi}\sqrt{\frac{|eB|}{2\pi}}e^{-|eB|x'^2/2}\hat{f}_{n_u-1}\left(x+\frac{q_y}{e_uB}\right)\hat{f}_{n_u}\left(x'+\frac{q_y}{e_uB}\right)\hat{f}_{n_d-1}\left(x'+\frac{q_y}{e_dB}\right)\hat{f}_{n_d}\left(x+\frac{q_y}{e_dB}\right),\\
&g_{4,n_u,n_d}=\int dx \int dx'\int\frac{dq_y}{2\pi}\sqrt{\frac{|eB|}{2\pi}}e^{-|eB|x'^2/2}\hat{f}_{n_u}\left(x+\frac{q_y}{e_uB}\right)\hat{f}_{n_u}\left(x'+\frac{q_y}{e_uB}\right)\hat{f}_{n_d}\left(x'+\frac{q_y}{e_dB}\right)\hat{f}_{n_d}\left(x+\frac{q_y}{e_dB}\right).
\end{aligned}
\end{equation}
Completing the integration, one obtains
\begin{equation}
\begin{aligned}
&g_{1,n_u,n_d}=|eB| \hat{e}_u^{n_d+1} \hat{e}_d^{n_u+1}\frac{\Gamma(n_u+n_d+1)}{\Gamma(n_u+1)\Gamma(n_d+1)},\\
&g_{2,n_u,n_d}=-|eB| \hat{e}_u^{n_d+3/2} \hat{e}_d^{n_u+3/2}\frac{\Gamma(n_u+n_d+2)}{\Gamma(n_u+1)\Gamma(n_d+1)}\sqrt{\frac{1}{(n_u+1)(n_d+1)}},\\
&g_{3,n_u,n_d}=g_{2,n_u-1,n_d-1},\\
&g_{4,n_u,n_d}=g_{1,n_u,n_d},
\end{aligned}
\end{equation}
where $\hat{e}_f=|e_f/e|$. Double summing with respect to $n_u$ and $n_d$, it gives that
\begin{equation}
\begin{aligned}
K_{EE}^{\pi_{+}}&=\int_{\tau_{uv}^2}^{\tau_{ir}^2} d\tau\int_{0}^{1}d\alpha\frac{|eB| \hat{e}_u \hat{e}_d e^{-\tau\omega_{ud}(\alpha,\bar{P}^2)}}{4\pi^2\tau}\left\{\frac{(e^{-2|eB| \hat{e}_u\hat{\alpha}\tau}+e^{-2|eB| \hat{e}_d \alpha\tau})(1+(M_uM_d-\alpha\hat{\alpha}\bar{P}^2-\omega_{ud}(\alpha,\bar{P}^2))\tau)}{1-\hat{e}_de^{-2|eB| \hat{e}_u\hat{\alpha}\tau}-\hat{e}_u e^{-2|eB| \hat{e}_d \alpha\tau}}\right.\\
&\quad\left.-\frac{2|eB| \hat{e}_d \alpha\tau e^{-2|eB| \hat{e}_d \alpha\tau}(1-(\hat{e}_u+\hat{e}_d)e^{-2|eB| \hat{e}_u\hat{\alpha}\tau})}{(1-\hat{e}_d e^{-2|eB| \hat{e}_u\hat{\alpha}\tau}-\hat{e}_u e^{-2|eB| \hat{e}_d \alpha\tau})^2}-\frac{2|eB| \hat{e}_u\hat{\alpha}\tau e^{-2|eB| \hat{e}_u\hat{\alpha}\tau}(1-(\hat{e}_u+\hat{e}_d)e^{-2|eB| \hat{e}_d \alpha\tau})}{(1-\hat{e}_d e^{-2|eB| \hat{e}_u\hat{\alpha}\tau}-\hat{e}_u e^{-2|eB| \hat{e}_d \alpha\tau})^2}\right\},\\
K_{EF}^{\pi_{+}}&=\int_{\tau_{uv}^2}^{\tau_{ir}^2} d\tau\int_{0}^{1}d\alpha\frac{|eB| \hat{e}_u \hat{e}_d e^{-\tau\omega_{ud}(\alpha,\bar{P}^2)}}{4\pi^2}\frac{\bar{P}^2(\alpha M_d+\hat{\alpha}M_u)(e^{-2|eB| \hat{e}_u\hat{\alpha}\tau}+e^{-2|eB| \hat{e}_d \alpha\tau})}{2M(1-\hat{e}_de^{-2|eB| \hat{e}_u\hat{\alpha}\tau}-\hat{e}_u e^{-2|eB| \hat{e}_d \alpha\tau})},\\
K_{FE}^{\pi_{+}}&=\frac{2M^2}{\bar{P^2}}K_{EF}^{\pi_{+}},\\
K_{FF}^{\pi_{+}}&=-\frac{M_u+M_d}{2M}K_{FE}^{\pi_{+}},
\end{aligned}
\end{equation}
where $\omega_{ud}(\alpha,\bar{P}^2)=\alpha M_d^2+\hat{\alpha}M_u^2+\alpha\hat{\alpha}\bar{P}^2$. The normalization of the BSA is determined by
\begin{eqnarray}
\left[\frac{\partial \ln(\lambda(\bar{P}^2))}{\partial \bar{P}^2}\right]^{-1}&=&-2N_c\sum_{X=E,F}\left(K_{EX}^{\pi_{+}}X_{\pi_{+}}E_{\pi_{+}}+\frac{\bar{P}^2}{2M^2}K_{FX}^{\pi_{+}}X_{\pi_{+}}F_{\pi_{+}}\right).
\end{eqnarray}
Moreover, the decay constants are given by
\begin{eqnarray}
f_{\pi}&&=\frac{N_c}{M}\left(K_{FE}^{\pi_{+}}E_{\pi_{+}}+K_{FF}^{\pi_{+}}F_{\pi_{+}}\right),
\end{eqnarray}
\begin{equation}
f'_{\pi}\left|eB\right|=\frac{N_c}{M}\left(\tilde{K}_{FE}^{\pi_{+}}E_{\pi_{+}}+\tilde{K}_{FF}^{\pi_{+}}F_{\pi_{+}}\right),
\end{equation}
where
\begin{eqnarray}
\tilde{K}_{FE}^{\pi_{+}}&&=\int_{\tau_{uv}^2}^{\tau_{ir}^2} d\tau\int_{0}^{1}d\alpha\frac{|eB| \hat{e}_u \hat{e}_d e^{-\tau\omega_{ud}(\alpha,\bar{P}^2)}}{4\pi^2}\frac{M(\alpha M_d+\hat{\alpha}M_u)(-e^{-2|eB| \hat{e}_u\hat{\alpha}\tau}+e^{-2|eB| \hat{e}_d \alpha\tau})}{(1-\hat{e}_de^{-2|eB| \hat{e}_u\hat{\alpha}\tau}-\hat{e}_u e^{-2|eB| \hat{e}_d \alpha\tau})},\\
\tilde{K}_{FF}^{\pi_{+}}&&=-\frac{M_u+M_d}{2M}\tilde{K}_{FE}^{\pi_{+}}.
\end{eqnarray}
%==================================================
%\bibliography{refclean}
%\end{document}
\providecommand{\noopsort}[1]{}\providecommand{\singleletter}[1]{#1}%

\end{document}